\documentclass[aps,twocolumn,superscriptaddress,nofootinbib,longbibliography]{revtex4-2}
\usepackage{dcolumn}
\usepackage{bm}
\usepackage{graphicx}
\usepackage{amsfonts}
\usepackage{bbm}
\usepackage{dsfont}
\usepackage{float}
\usepackage{graphicx}
\usepackage{bbm}
\usepackage{amssymb}
\usepackage{multibib}
\usepackage{color}
\usepackage{enumerate}
\usepackage{amsmath}
\usepackage{amstext}
\usepackage{latexsym}
\usepackage{braket}
\usepackage{appendix}
\usepackage{textcomp}
\usepackage[usenames,dvipsnames]{xcolor}
\usepackage[colorlinks=true,citecolor=blue,linkcolor=RubineRed,urlcolor=blue]{hyperref}
\usepackage{mathtools}
\usepackage{lipsum}
\usepackage{grffile}
\usepackage{subfigure}
\usepackage{overpic}

\begin{document}
\title{Universal Vortex Statistics and Stochastic Geometry 
of Bose-Einstein Condensation}

\author{Mithun Thudiyangal\href{https://orcid.org/0000-0003-4341-6439}{\includegraphics[scale=0.05]{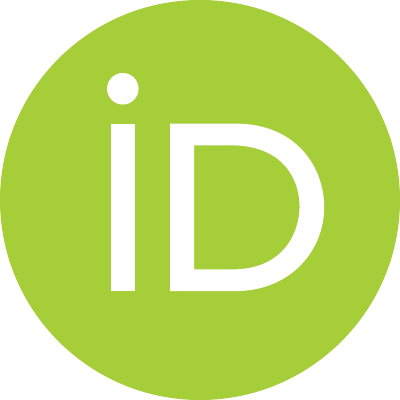}}}
\affiliation{Department  of  Atomic and Molecular Physics,  Manipal Academy of Higher education, Manipal 576 104, India}
\affiliation{Department  of  Physics  and  Materials  Science,  University  of  Luxembourg,  L-1511  Luxembourg, G. D.  Luxembourg}
\author{Adolfo del Campo\href{https://orcid.org/0000-0003-2219-2851}{\includegraphics[scale=0.05]{orcidid.pdf}}}
\affiliation{Department  of  Physics  and  Materials  Science,  University  of  Luxembourg,  L-1511  Luxembourg, G. D.  Luxembourg}
\affiliation{Donostia International Physics Center,  E-20018 San Sebasti\'an, Spain}

\begin{abstract}
The cooling of a Bose gas in finite time results in the formation of a Bose-Einstein condensate that is spontaneously proliferated with vortices. We propose that the vortex spatial statistics is described by a homogeneous Poisson point process (PPP) with a density dictated by the Kibble-Zurek mechanism (KZM). We validate this model using numerical simulations of the two-dimensional stochastic Gross-Pitaevskii equation (SGPE) for both a homogeneous and a hard-wall trapped condensate. The KZM scaling of the average vortex number with the cooling rate is established along with the universal character of the vortex number distribution. The spatial statistics between vortices is characterized by analyzing the two-point defect-defect correlation function, the corresponding spacing distributions, and the random tessellation of the vortex pattern using the Voronoi cell area statistics. Combining the PPP description with the KZM, we derive universal theoretical predictions for each of these quantities and find them in agreement with the SGPE simulations. Our results establish the universal character of the spatial statistics of point-like topological defects generated during a continuous phase transition and the associated stochastic geometry.

\end{abstract}
\maketitle

Spontaneous symmetry breaking in finite time leads to the formation of topological defects. A paradigmatic example concerns cooling a Bose gas below the temperature for Bose-Einstein condensation. This is characterized by a continuous phase transition signaled by the growth of the condensed fraction that acts as the order parameter \cite{Ueda10}. The formation of a Bose-Einstein condensate (BEC) involves the breaking of $U(1)$ symmetry and, in a pancake atomic cloud, leads to the proliferation of spontaneously-formed vortices \cite{Weiler08}, topological defects characterized by an integer-valued winding number $w\in\mathbb{Z}$. In this context, vortices with $|w|>1$ are unstable against the decay into vortices with $w=\pm 1$ of lower energy.

In ultracold atoms, the density of vortices can be directly measured, e.g., by absorption imaging after a time of flight \cite{Ketterle01,Anderson10} or by in situ imaging \cite{Wilson15}. In addition, the characterization of vortex patterns in BEC can be automatized using deep learning algorithms \cite{Metz21,Kim23}. This has promoted the use of ultracold gases as a test bed for nonequilibrium statistical mechanics. In particular, it has made it possible to probe the validity of the Kibble-Zurek mechanism (KZM), a universal paradigm describing the nonequilibrium dynamics across a continuous phase transition in finite time, which predicts the formation of topological defects  \cite{Kibble76a,Kibble76b,Zurek96a,Zurek96b,DZ14}.   
To this end, KZM relies on the equilibrium scaling relations for the correlation length $\xi$ and the relaxation time $\tau$ as a function of the proximity to the critical point $\varepsilon=(\lambda_c-\lambda)/\lambda_c$. Here, $\lambda$ is the control parameter driving the transition at the critical point $\lambda_c$. Both $\xi$ and $\tau$ obey universal power laws
\begin{eqnarray}
\xi(\varepsilon)&=&\frac{\xi_0}{|\varepsilon|^{\nu}},\label{xieq}\\    \tau(\varepsilon)&=&\frac{\tau_0}{|\varepsilon|^{z\nu}}\label{taueq},
\end{eqnarray}
where $\nu$ and $z$ are critical exponents and $\xi_0$ and $\tau_0$ are system-dependent  constants.
A finite time quench of the control parameter $\lambda(t)= \lambda_c(1-t/\tau_Q)$ leading to the linear variation of $\varepsilon(t)=t/\tau_Q$ sets a universal scaling of nonequilibrium properties. The KZM identifies the typical response time in which the order parameter starts to grow after crossing the critical point.  
This time scale, often referred to as the freeze-out time, is derived by matching the time elapsed after crossing the critical point with the instantaneous relaxation time, $t=\tau_0/|t/\tau_Q|^{z\nu}$, and reads
\begin{eqnarray}
\hat{t}=\left(\tau_0\tau_Q^{z\nu} \right)^{\frac{1}{1+z\nu}}.
\label{tfreezeout}
\end{eqnarray}
Note the universal power-law scaling with the quench time $\tau_Q$, with an exponent fixed by the critical exponents of the relevant universality class characterizing the transition.  
The freeze-out time $\hat{t}$ further identifies a characteristic value of the control parameter away from the critical point in which the system responds to the quench, in dimensionless form  $\varepsilon(\hat{t})=(\tau_0/\tau_Q)^{\frac{1}{1+z\nu}}$,  and 
the correlation length out of equilibrium
\begin{equation}
\hat{\xi}=\xi[\varepsilon(\hat{t})]=\xi_0   \left(\frac{\tau_{Q}}{\tau_0}\right)^{\frac{ \nu}{1+z\nu}}.
\label{KZMxi}
\end{equation}
This prediction sets the density of point-like topological defects, which scales as
\begin{equation}
\rho 
=
\frac{1}{\xi_0^d}\left(\frac{\tau_0}{\tau_{Q}}\right)^{\frac{d \nu}{1+z\nu}},
\label{rhoKZM}
\end{equation}
in $d$-spatial dimensions. 
For a pancake BEC, effectively $d=2$ and one expects the vortex density to scale with the cooling time $\tau_Q$ as $\rho\propto \tau_Q^{-\frac{2\nu}{1+z\nu}}$.
In the mean-field regime characterized by $\nu = 1/2$ and $z = 2$, KZM  predicts the vortex density to scales as $\rho\propto \tau_Q^{-1/2}$ in a spatially-homogeneous system.
However, ultracold gases in the laboratory do not comply with mean-field behavior. The experimental study of the equilibrium scaling relation for the correlation length (\ref{xieq}) in the BEC transition \cite{Donner07}  reveals a value of $\nu$ closely matching that of the 3D XY universality class, $\nu=0.6717(1)$ \cite{Campostrini01}. 
The KZM prediction for the nonequilibrium correlation length (\ref{KZMxi}) has been investigated by Navon et al. \cite{Navon15}, who reported experimental measurements consistent with the value of the dynamical critical exponent $z=3/2$ expected for this universality class \cite{Hohenberg_1977}.
 The vortex density scaling predicted by the KZM has been experimentally demonstrated in a Bose gas \cite{Chomaz15,Donadello16,Goo21,Rabga23} and a strongly interacting Fermi gas \cite{Shin19,lee2023observation}.

The ubiquitous confinement potential in experiments with trapped ultracold gases can induce a spatial dependence of the critical point and make the transition inhomogeneous  \cite{KV97,Zurek09,DRP11,Nikoghosyan16}. The role of causality in the course of symmetry breaking is then enhanced by restricting the formation of topological defects and can lead to a steepening of the KZM scaling as predicted by the so-called Inhomogeneous KZM \cite{DKZ13}, studied theoretically in the context of vortex formation in BEC \cite{DRP11} and observed in experiments with trapped ions \cite{Ulm13,Pyka13} and Bose gases  \cite{KimShin22,Rabga23}. However, such deviations from the canonical KZM scaling can be suppressed by resorting to fast quenches \cite{delcampo10,DRP11,DKZ13}, below the threshold rate for defect saturation \cite{Chesler:2014gya,Goo21,Zeng23}, or by using homogeneous trapping potentials,  such as ring and box-like traps \cite{Gupta05,Henderson09,Gaunt13,Navon15,Chomaz15,Navon16,Mukherjee17,Guillaume2019giant}.

The recent quest for signatures of universality in the critical dynamics beyond the KZM has focused on the number distribution of topological defects \cite{GomezRuiz20,Mayo21,delcampo21,Goo21,Subires22,GomezRuiz22}. The number of spontaneously formed topological defects fluctuates in different experimental runs, as well as  numerically simulated realizations. The probability of observing a given number of defects is well described by a  binomial distribution, in which all cumulants follow the same power-law scaling with the quench rate \cite{GomezRuiz20,Mayo21,delcampo21,Subires22,GomezRuiz22}. 

 Despite their success, the KZM and its generalizations leave without answering aspects regarding the spatial distribution of topological defects. The latter is of key importance in a variety of applications ranging from condensed matter and nanotechnology (probing, characterizing, and controlling defects and their interactions) to the study of turbulence, structure formation, and material aging, to name some examples.

This manuscript establishes the universal signatures of the spatial statistics of spontaneously formed topological defects. To this end, we combine the nonequilibrium scaling theory of phase transitions with tools of spatial statistics and stochastic geometry, the branch of probability theory concerned with random spatial patterns \cite{Chiu2013stochastic}. We model spatial correlations of vortices formed during Bose-Einstein condensation using a homogeneous Poisson Point Process (PPP) on a plane with a density dictated by  KZM, i.e., Eq. (\ref{rhoKZM}). We characterize the vortex-spacing distribution with and without conditioning on the winding number $w$. We further analyze the two-point correlations by mapping them to the celebrated Disk Line Picking problem in geometric probability.
In addition, we characterize the emergent stochastic geometry of the spontaneously formed vortices. Specifically, we analyze the Voronoi area cell distribution in a random tesselation of the vortex pattern. Universal predictions for these quantities derived from the PPP-KZM  model accurately reproduce numerical simulations of the Bose-Einstein condensation using the stochastic Gross-Pitaevskii equation. Our findings establish the universal spatial statistics and stochastic geometry of topological defects generated via the KZM.

\section{Kibble-Zurek Dynamics of the BEC Transition}
The formation of a BEC is signaled by the emergence of a nonzero complex order parameter $\Psi=\Psi(\mathbf{r},t)$, known as the condensate wavefunction.
We consider the two-dimensional (2D) stochastic Gross-Pitaevskii equation \cite{Hohenberg_1977,Damski_soliton_2010,Liu_kibble_2020,cockburn2009stochastic,blakie2008dynamics,stoof1999coherent,stoof2001dynamics} as a test bed for the spatial statistics of topological defects formed in a 3D pancake BEC. Specifically, we consider the evolution of the condensate wavefunction $\Psi=\Psi(\mathbf{r},t)$  according to 
\begin{equation}
(i-\gamma)\frac{\partial \Psi }{\partial t}=\Big[-\frac{1}{2}\nabla^2+g |\Psi |^{2} +V(\mathbf{r})+\varepsilon(t) \Big]\Psi+\xi(\mathbf{r},t),  \label{eq:lg1}
\end{equation} 
where $V(\mathbf{r})$ represents the external trap and $\gamma$ is the dissipation rate \cite{Choi:1998phenomenological}. The white noise $\xi$ is a complex Gaussian process with zero mean. It  satisfies the fluctuation-dissipation theorem
\begin{equation}
\langle\xi^{\ast}(\mathbf{r},t)\xi(\mathbf{r}',t')\rangle = 2 \gamma T \delta(\mathbf{r}-\mathbf{r}')\delta(t-t'),  \label{eq:lg2}
\end{equation}%
where $T$ is the temperature of the thermal cloud. 
 The physical quantities in Eq.~\eqref{eq:lg1} are dimensionless. The units of length, time and energy are respectively given by $\xi$, $m \xi^2/\hbar$ and $\hbar^2/(m\xi^2)$ and are defined in terms of the healing length $\xi=\sqrt{\hbar^2/(mgn_0)}$, where $m$ is the atomic mass and $n_0$ is the density of the uniform condensate ($V=0$). Additionally, temperature is measured in $\hbar^2/(k_Bm\xi^2)$ units, where $k_B$ is the Boltzmann constant. 
The equation of motion (\ref{eq:lg1}) conserves the norm,  
\begin{equation}
\mathcal{N}=\int_{-\infty }^{+\infty }n(\mathbf{r})~d\mathbf{r}=1,~~~\text{and}~~~n(\mathbf{r})=|\Psi (\mathbf{r})|^{2}, \label{eq2}
\end{equation}%
for $\gamma=0$. In the absence of an external trap ($V=0$), Eq. (\ref{eq:lg1}) is associated with the effective potential $V_e(\Psi)=\frac{g}{2}\Psi^4+\varepsilon \Psi^2$. The control parameter $\varepsilon(t)=-\mu(t)$ is set by the chemical potential $\mu$. Its variation across the critical point
 $\varepsilon_c=0$ describes the formation of a BEC, bringing
the system from a symmetric phase with $\Psi =0$ ($\varepsilon<0$) to a symmetry-broken phase represented by a ground-state complex wave function $\Psi =\sqrt{|\mu|/g}\exp(i\theta)$ ($\varepsilon>0$), where $\theta$ represents the global phase of the condensate. 
In the nonequilibrium setting, the polar decomposition of the condensate wavefunction reads $\Psi(\mathbf{r},t)=|\Psi(\mathbf{r},t)|\exp[i\theta(\mathbf{r},t)]$. The condensate velocity $v(\mathbf{r},t)=\frac{\hbar}{m}\nabla \theta(\mathbf{r},t)$  may acquire a net circulation around a given point 
\begin{eqnarray}
\oint v(\mathbf{r},t)\cdot d\ell=\frac{2\pi\hbar}{m}w,\quad w\in\mathbb{Z}.
\end{eqnarray}
The associated singularity constitutes a vortex, which is characterized by a vanishing density at its core and a finite healing length.
According to KZM, vortices spontaneously form during the phase transition.
To describe the critical dynamics of BEC formation, we consider a linear quench of the control parameter 
\begin{equation}
\varepsilon(t)=-\mu_i-\frac{t}{\tau_Q}|\mu_f-\mu_i|, 
\end{equation}
where $\tau_Q$ is the quench time, $\mu_i$ is the initial chemical potential, and $\mu_f$ is the final chemical potential.  To solve  Eq.~\eqref{eq:lg1} numerically, we use the software package XMDS \cite{dennis2013xmds2}. This software efficiently solves stochastic differential equations numerically by using a semi-implicit algorithm. Unless otherwise specified, we consider a homogeneous condensate, $V(\mathbf{r})=0$, with the domain size $\mathcal{D}=4 \times L^2$, where $L=15$. We further fix the parameters $\mu_i = 0.1$, $\mu_f = 20$, $g=1$, $\gamma=0.03$ and $T=10^{-6}$. Additionally,  we build an ensemble out of  $\mathcal{R}$ stochastic trajectories corresponding to different noise realizations to characterize the spatial statistics. We begin our numerical experiments with the initial condition $\Psi=0$ and let the system relax 
for a time $t_{0}=10$ before the beginning of a quench. 


    \begin{figure}[t]
  \includegraphics[width=1\columnwidth,angle =0]{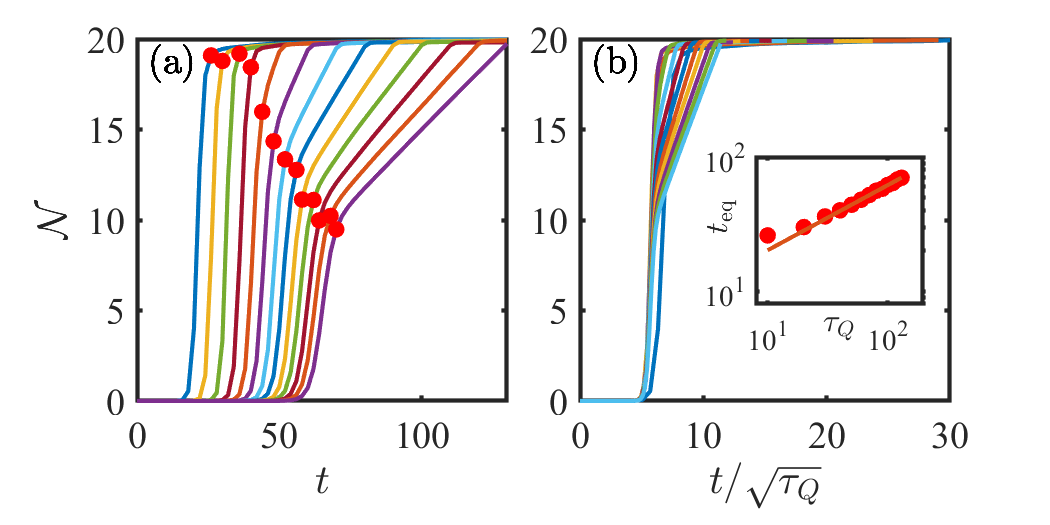}
       \caption{Kibble-Zurek universality in the response time of the order parameter in a BEC transition. a) Norm $\mathcal{N}$ as a function of the time of evolution following a linear ramp of the chemical potential with $\tau_Q=(10,20,...,130)$ (left to right) in a single realization. Red points indicate the crossover from an exponential growth to a linear growth, and the corresponding time is denoted by $t_{\text{eq}}$. (b) The collapse of the growth of the BEC quantified by $\mathcal{N}$ as a function of the evolution time scaled by the freeze-out time $\hat{t}\propto\sqrt{\tau_Q}$.  For slow quenches ($\tau_Q > 20$), all the lines collapse to a single line for $t<t_{\text{eq}}$. The inset in panel (b) shows  $t_{\text{eq}}$ as a function of $\tau_Q$ with a scaling $t_{\text{eq}} \propto \tau_Q^{0.48\pm0.046}$ indicated by the solid line. The scaling of $t_{\text{eq}}$ is consistent with the KZM prediction for $\hat{t}$ in Eq. (\ref{tfreezeout}). 
       }
    \label{fig:tqall}
\end{figure}

\section{Universal vortex statistics}
Above the BEC transition, the order parameter effectively vanishes. During the phase transition, it grows in a time scale set by the freeze-out time $\hat{t}$.
KZM introduces the nonequilibrium correlation length scale $\hat{\xi}$. We model the vortex spatial statistics by assuming that a mosaic of protodomains forms with an average length scale set by $\hat{\xi}$ and tessellate the atomic cloud in the early stages of the dynamics. In each domain, we assume the phase of the emergent BEC to be homogeneous and chosen at random. When such domains merge at a point, a vortex is formed with probability $p$ \cite{Scherer07,YatesPRL98}.

     \begin{figure*}
\includegraphics[width=1.56\columnwidth,angle =0]{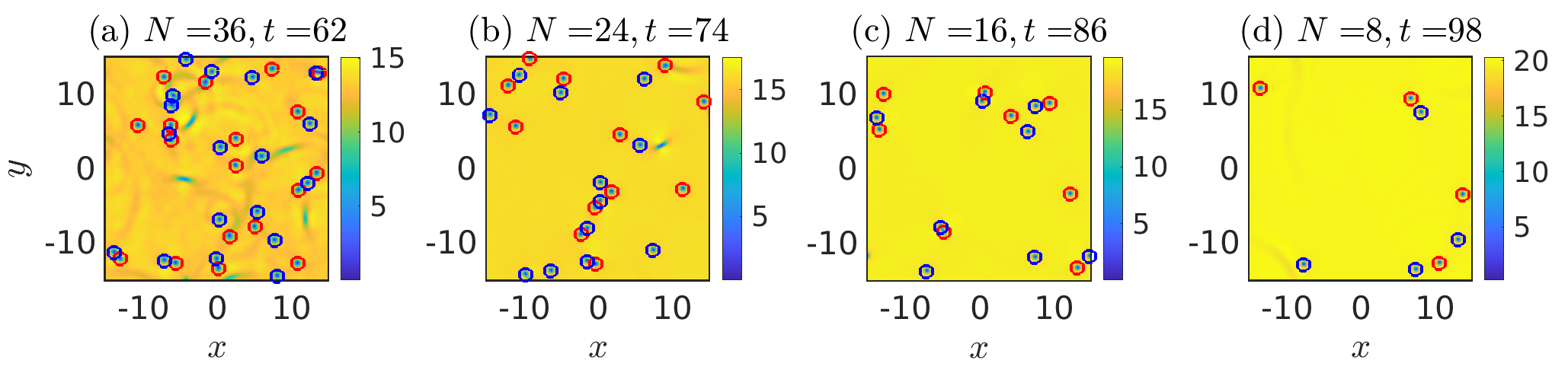}
    \includegraphics[width=0.48\columnwidth,angle =0]{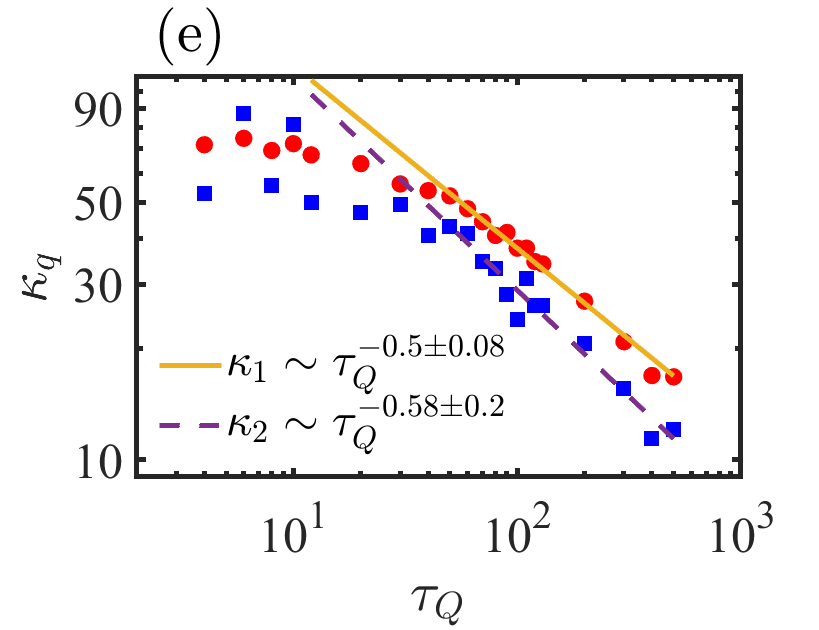}
       \caption{Spontaneous vortex formation resulting from BEC formation. Panels (a)-(d) show snapshots of the nonequilibrium condensate density $|\Psi|^2$ with spontaneously formed vortices 
       in a homogeneous 2D BEC at different times  ($t > t_{\text{eq}}$) for a fixed quench time $\tau_Q=90$. The total vortex number $N$ is the sum of vortices with positive and negative charges, represented by red and blue circles, respectively. The total topological charge is zero in each realization. (e) Dependence of the total vortex number $N=\kappa_1$ on the quench time $\tau_Q$, determined at the time of evolution $t=t_{\text{eq}}$. The red symbols show the average over $\mathcal{R}=400$ different runs. 
       As the quench time is increased the total vortex number exhibits a crossover from a plateau value, observed in the limit of fast quenches, to a power-law behavior. The fluctuations of the vortex number share this behavior.  
         The fits to the first two cumulants, the average ($\kappa_1$) and variance ($\kappa_2$) of the total number of defects $N$,
      agree with the universal cumulant power-law prediction in Eq. (\ref{eq:kqFCS}) with mean-field critical exponents  $\nu = 1/2$ and $z = 2$.
       }
    \label{fig:densitytq90}
\end{figure*}

\subsection{Universal BEC growth dynamics}
Figure \ref{fig:tqall}a shows the dynamics of BEC formation as monitored by the growth of the condensate wavefunction norm when the chemical potential is varied in different quench time scales. According to the KZM, the response time of the nonequilibrium system is set by the freeze-out time $\hat{t}=(\tau_0\tau_Q^{z\nu})^{\frac{1}{1+z\nu}}$. This universal scaling can be revealed by monitoring the growth of the order parameter  in real time for different values of the quench rate and rescaling the time of evolution by the freeze-out time $\hat{t}$ \cite{Damski10,Sonner15,Reichhardt22}. This prediction is verified in Fig. \ref{fig:tqall}b, which leads to a collapse of the different curves of the left panel. In addition, numerical simulations show that for a given quench time $\tau_Q$, the growth of the condensate norm exhibits a transition from an early exponential regime to a later stage characterized by a linearly-in-time behavior. We refer to the crossover time as the equilibration time $t_{\text{eq}}$ (marked by red points), which is expected to be proportional to $\hat{t}$ and thus scales with the quench time as $t_{\text{eq}}\propto\sqrt{\tau_Q}$ when $z\nu=1$. 
An analogous time scale has been identified in the holographic superfluids \cite{Chesler:2014gya}. 
Its scaling with the quench time is verified in the inset of the right panel. For fast quenches, $\hat{t}$ deviates from the KZM scaling  $\sqrt{\tau_Q}$, saturating at a plateau where the vortex number becomes independent of the quench time. This behavior agrees with the breakdown of the KZM in the fast quench limit  \cite{delcampo10,Chesler:2014gya,Donadello16,Zeng23}. We focus on the range of quench times in which KZM holds. In this regime, the time scale in which the BEC is formed exhibits a universal power-law scaling as a function of the quench time in which the transition is driven.

At the crossover time $t_{\text{eq}}$, the expected number of defects is maximum due to the development of well-defined vortex cores resulting from merging multiple protodomains. The exponential and linear stages of the BEC growth under a slow quench are associated with regimes of vortex generation and adiabatic BEC growth, respectively \cite{Chesler:2014gya}. Fig.~\ref{fig:densitytq90} shows the condensate density $|\psi|^2$ with both positively and negatively charged vortices at different times of the dynamical process for $t\sim t_{\text{eq}}$. In addition, our simulations of the homogeneous BEC formation rely on the SGPE with periodic boundary conditions. As a result of the Poincar\'e-Hopf theorem, the total topological charge of the system at any given time should vanish identically. This is  verified in Fig.~\ref{fig:densitytq90}. 
Further, the number of vortices following the quench generally decreases with the passage of time due to vortex antivortex annihilation by emitting sound waves.

\subsection{Vortex Number Statistics}
While our primary interest lies in the vortex spatial statistics and the associated stochastic geometry, we first characterize the fluctuations in the total number of spontaneously formed vortices. These fluctuations among independent realizations affect the spatial statistics and lie beyond the KZM.
A description of the vortex full counting statistics relies on the assumption that vortex formation at different locations is described by independent events \cite{GomezRuiz20,Mayo21}. This model uses the KZM correlation length to partition the emergent BEC into protodomains, in which the condensate phase is coherent. A vortex forms at the merging point between different domains with a given success probability according to the geodesic rule. Events for defect formation are assumed to be identically distributed, with a success probability $p$.
 The number of possible domain locations for defect formation can be estimated as $\mathcal{N}_d=A/(\hat{\xi}^2)$, for an emergent BEC of area $A$. The probability of forming $N$ defects is then given by the binomial distribution
\begin{eqnarray}
P(N)=
\binom{\mathcal{N}_d}{N}p^N (1-p)^{\mathcal{N}_d-N}.
\label{eq:binomial}
\end{eqnarray}
 This distribution encodes the universal scaling of its cumulants $\kappa_q$ with the  quench time $\tau_{Q}$,
 \begin{eqnarray}
\kappa_q \propto \left(\frac{\tau_0}{\tau_{Q}}\right)^{\frac{2 \nu}{1+z\nu}}.
\label{eq:kqFCS}
\end{eqnarray}

The distribution given in Eq.~\eqref{eq:binomial} approaches a normal distribution in the large $\mathcal{N}_d$ limit,
\begin{eqnarray} 
P(N)=\frac{1}{\sqrt{2\pi (1-p) \langle N \rangle}}\exp\left(-\frac{(N-\langle N \rangle)^2 }{2(1-p)\langle N \rangle}\right),
\label{eq:binomialLN}
\end{eqnarray}
where $\langle N \rangle$ represents the mean defect numbers averaged over different initial conditions.
 
 Fig.~\ref{fig:densitytq90}(e) shows the average ($k_1$) and variance ($k_2$) of the total defect number $N$ as a function of $\tau_Q$. In the slow quench regime, the first two cumulants $\kappa_1$ and $\kappa_2$ follow the scaling law $\kappa_1 \propto\tau_Q^{-0.48 \pm 0.08 }$ and $\kappa_2 \propto\tau_Q^{-0.58 \pm 0.2 }$, respectively.  These scaling are in good agreement with the beyond-KZM prediction in Eq.~\eqref{eq:kqFCS}, where for mean-field critical exponents $\frac{2 \nu}{1+z\nu}=1/2$.


\subsection{Vortex Spatial Statistics and Stochastic Geometry}
We model the location of the vortices using a homogeneous Poisson point process with a density determined by the KZM scaling law, i.e., the PPP-KZM model. Building on it, we next derived closed-form distributions characterizing spontaneous vortex patterns and the spatial correlation between vortices.

\subsubsection{Vortex Distance Distribution}
We first characterize the distribution of the distance between any two vortices. Vortices are assumed to be randomly distributed on a disk of radius $R$. Given the vortex locations $\mathbf{r}$ and $\mathbf{r}'$, we consider the distance $s=\|\mathbf{r}-\mathbf{r}'\|$.

The distribution of the distance $s$  between vortices is then that of the Disk Line Picking problem in geometric probability  \cite{mathai1999introduction}.
The vortex distance distribution is 
\begin{eqnarray}
 P_2(s)=\frac{4s}{\pi R^2}\bigg[\arccos\left(\frac{s}{2R}\right)-\frac{s}{2 R} \sqrt{1-\frac{s^2}{4R^2}}\bigg],
 \label{eq:dist2}
\end{eqnarray}
with the mean, $\langle s \rangle =\int_0^{2 R} sP(s)ds= r R,~~r=\frac{128}{45\pi} $. A crucial feature of $P_2(s)$ is that it is ``blind'' to the KZM scaling, as it is independent of the total number of vortices. As long as the vortex locations are described by a PPP, the result for $P_2(s)$ is the same whether it is computed for a pair of vortices or a hundred. As such, $P_2(s)$ is ideally suited to test the validity of the PPP description, independently of the validity of the KZM scaling.

\subsubsection{Unconditioned Vortex Spacing Distribution}
Given the location of a vortex as a reference, the spacing distribution is defined as the probability of finding any of the other $(N-1)$ vortices at a distance between $s$ and $s + ds$, with the remaining $(N-2)$ vortices being located farther away.   This definition treats vortices with winding number $w=\pm 1$ on equal footing, without distinction by their topological charge. 

\begin{figure}
\includegraphics[width=0.38\columnwidth,angle =0]{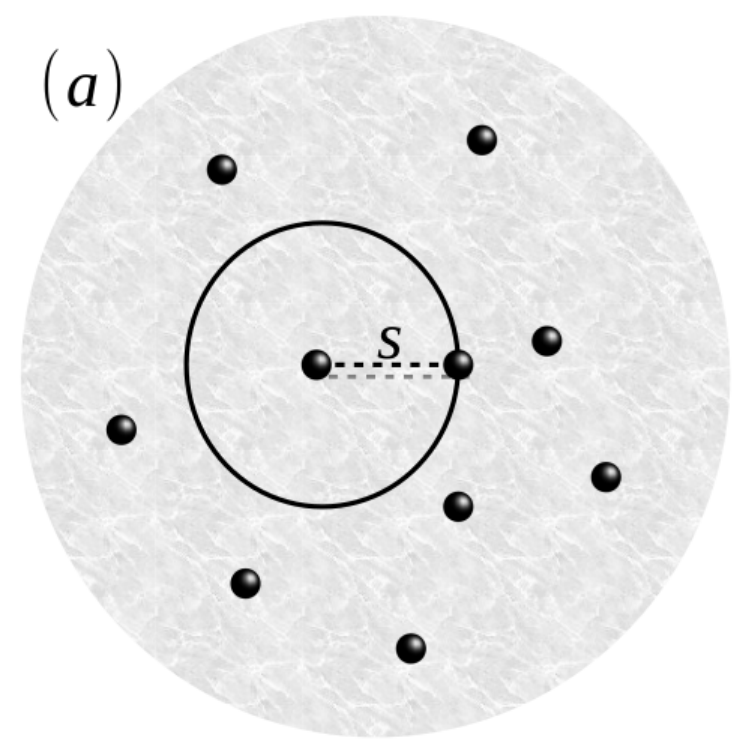}
\includegraphics[width=0.38\columnwidth,angle =0]{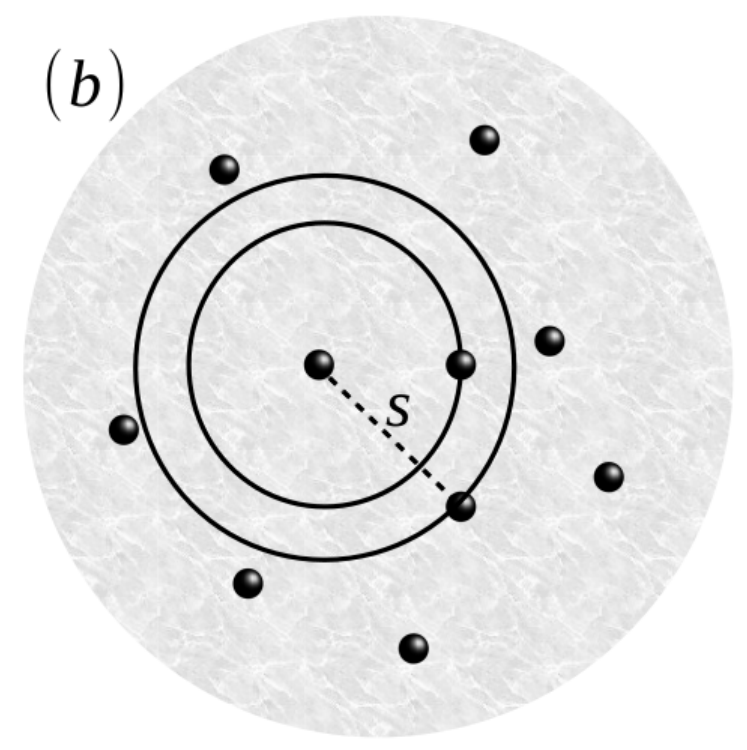}
       \caption{Schematic representation of the defect spacing $s$ with respect to the reference vortex at the center of the circle in the case of (a) $P(s)=P^{(1)}(s)$  and (b) $P^{(2)}(s)$.  The black points represent vortices without accounting for their topological charges.
       }
    \label{fig:cartoon1}
\end{figure}

A schematic representation of the first nearest neighbor distance $s$ is shown in Fig.~\ref{fig:cartoon1}. In the limit of large $N$, and using the spacing normalized to the mean $S=s/\langle s\rangle$ the spacing distribution takes the form of a Wigner-Dyson distribution \cite{delcampo22}
\begin{eqnarray}
P^{(1)}(S)=P(S)=\frac{\pi}{2}S\exp\left(-\frac{\pi}{4}S^2\right).
\label{eq:poissonian}
\end{eqnarray}
In this expression, the KZM universality is hidden in the scaling of the mean spacing, predicted to be
\begin{eqnarray}
\langle s\rangle=\frac{\sqrt{\pi}}{2}\hat{\xi}\propto \left(\frac{\tau_{Q}}{\tau_0}\right)^{\frac{ \nu}{1+z\nu}}.
\label{eq:means}
\end{eqnarray}

\subsubsection{Vortex Spacing Distribution of $k$-th Order}
The spacing distribution in Eq.~\eqref{eq:poissonian} can be generalized by focusing on the $k$-th nearest neighbor; see Fig.~\ref{fig:cartoon1} for a  schematic representation. To be precise, considering a vortex of reference at the origin, with its first  $(k-1)$ neighbors being located in the interval $(0,s)$, the distribution $P^{(k)}(s)$ describes the probability to find the $k$-th neighbor at a distance between $s$ and $s+ds$, provided that the remaining $N-k-1$ vortices are farther apart.  
An explicit computation using the PPP-KZM model shows that for the spacing relative to the mean, 
\begin{eqnarray}
P^{(k)}(S)=  \frac{2}{(k-1)\,!} r^{2k-1}S^{2k-1}e^{- r^2 S^2},
\label{eq:kneighborff}
\end{eqnarray}
where $r=\frac{\Gamma \big(\frac{1}{2}+k \big)}{\Gamma (k)}$, as detailed in  Appendix~\ref{app:ppmmkth}. This is in agreement with the known $k$-th order spacing distribution of a PPP with unit mean spacing \cite{mathai1999introduction}.

\begin{figure}[t]
 \includegraphics[width=1\columnwidth,angle =0]{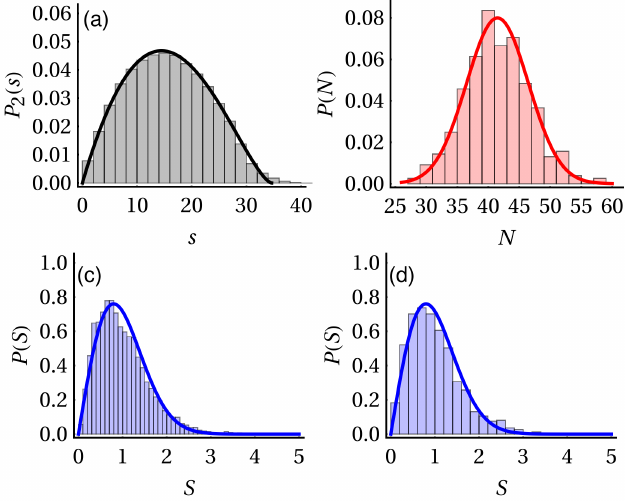}
       \caption{Histograms of (a) the distance $s$ between two defects, (b) the total number of defects $N$, and (c) the  spacing between topological defects $S$ relative to the mean, for varying $N$, (d) and for fixed $N$ (bottom panel) at $t=t_{\text{eq}}$ for $\tau_Q=90$.
       The solid black line in (a) represents the Disk Line Picking distribution Eq.~\eqref{eq:dist2}, the solid red line in (b) denotes the normal distribution Eq.~\eqref{eq:binomialLN}, and in panels (c)-(d), the solid blue line stands for the Wigner-Dyson distribution in Eq. \eqref{eq:poissonian}. 
       }
    \label{fig:pdf_nd_teq}
\end{figure}

Using these figures of merit, let us assess the validity of the PPP-KZM model in describing the vortex patterns in a newborn BEC.
The randomness of the numerically obtained vortex locations is verified in Fig.~\ref{fig:pdf_nd_teq}a. The distribution of the distance between two defects agrees well with the prediction in Eq.~\eqref{eq:dist2} for the distance between two random points on a disk, i.e., the Disk Line Picking problem  \cite{lellouche2019distribution}. The deviation in the tail of the probability distribution from the analytical prediction is attributed to the square geometry of the domain. 

As a result of thermal fluctuations, the vortex number is a stochastic variable, fluctuating among different realizations. The vortex number statistics follows the normal distribution Eq.~\eqref{eq:binomialLN}, a limiting case of the binomial distribution Eq.~\eqref{eq:binomial}. 

\begin{figure}
\includegraphics[width=1\columnwidth,angle =0]{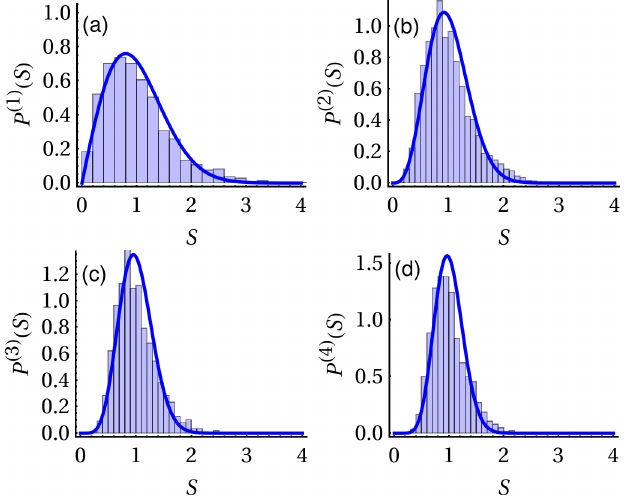}\\
       \caption{
       Histogram $P^{(k)}(S)$ of the spacing between topological defects $S$ for the  $k$-th nearest neighbor, with $k=1,2,3,4$ from left to right at $t=t_{\text{eq}}$,  and $\tau_Q=90$.
       The solid blue line in each panel represents the corresponding distribution in Eq.~\eqref{eq:kneighborff}.
       }
    \label{fig:pdf_higher}
\end{figure}

Given this, we analyze the spatial statistics for a fixed $N$ and for varying $N$. The corresponding histograms are shown in  Fig.~\ref{fig:pdf_nd_teq}(c)-(d) and are in good agreement with the Wigner-Dyson distribution Eq.~\eqref{eq:poissonian}. The difference in the spatial statistics with fixed and varying $N$ is negligible in the scaled spacing variable $S$.  
This is further verified and shown in the Appendix \ref{app:ppp} by considering a PPP of random points generated on a domain of size $\mathcal{D}=4 \times L^2$, where $L=15$. Additionally, as shown in Fig.~\ref{fig:pdf_higher}, an excellent agreement is found between the numerically calculated and analytically estimated probability distributions for the $k$-th nearest neighbor distance. We further verify the matching between numerical and analytical results in the Appendix \ref{app:confidenceband}.

Figure ~\ref{fig:defectscumulants} shows the mean spacing $\langle s \rangle$ calculated at $t=t_{\text{eq}}$ as a function of $\tau_Q$. 
At fast quenches, the average spacing remains constant. The vortex number saturates at a plateau in this limit due to the breakdown of the KZM scaling, recently shown to be universal \cite{Chesler:2014gya,Zeng23}.
For slower quench times, the mean spacing scales with $\tau_{Q}$ as $\langle s \rangle \propto \tau_{Q}^{ 0.27\pm 0.04}$.  This power law agrees with the prediction in Eq.~\eqref{eq:means} beyond the KZM. 
Similar statistics are found  at different times for $t\sim\mathcal{O}(t_{\text{eq}})$, as long as vortex decay arising from coarsening remains subdominant.

\subsubsection{Conditioned Vortex Spacing Distribution}
The PPP-KZM model assumes the location of vortices to be uncorrelated. In particular, it ignores the topological charge of the vortices. It also ignores that once fully formed, vortices interact with each other through an effective logarithmic potential as a two-dimensional Coulomb plasma \cite{Nelson02book}. As a result, vortices with equal topological charges repel each other, while those with opposite charges attract. In short, there are vortex-vortex and vortex-antivortex correlations that the PPP-KZM model ignores.

\begin{figure}[t]
\includegraphics[width=0.98\columnwidth,angle =0]{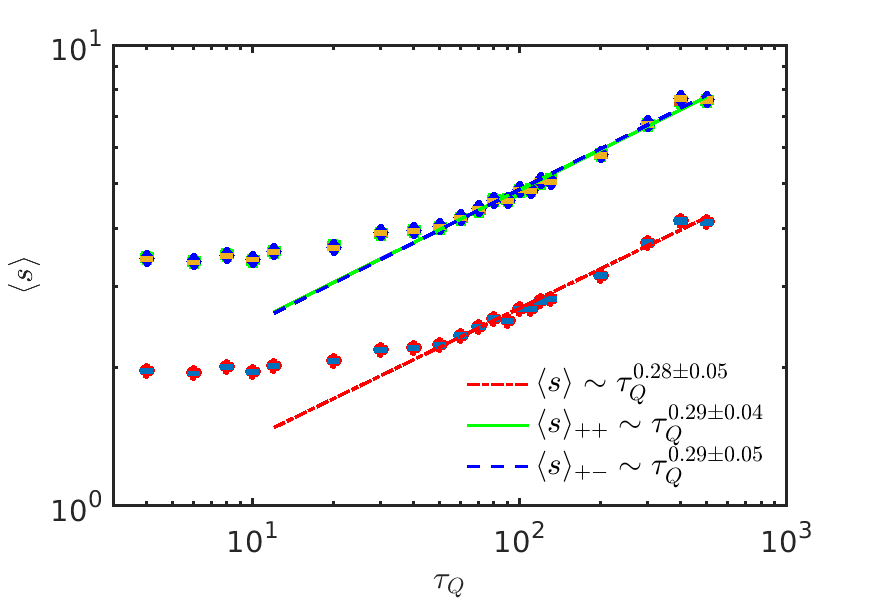}
        \caption{The mean spacing $\langle s \rangle$, $\langle s \rangle_{++}$ and $\langle s \rangle_{+-}$ are shown in red symbols as a function of $\tau_Q$. The values are computed at $t=t_{\text{eq}}$, using an average over $\mathcal{R}=400$ different runs, that sets the width of the error bars. 
        }
    \label{fig:defectscumulants}
\end{figure}

Given the restriction in the winding number $w=\pm1$, the total number of vortices is the sum of the number $N_+$ of those with $w=+1$ and the number $N_-$ of those with $w=-1$, i.e., $N=\sum_{w=\pm1}N_w=N_++N_-$.  
The effect of the interactions between vortices can be studied by introducing the generalization of the spacing distribution
conditioned on the topological charge.

\begin{figure}
\includegraphics[width=0.38\columnwidth,angle =0]{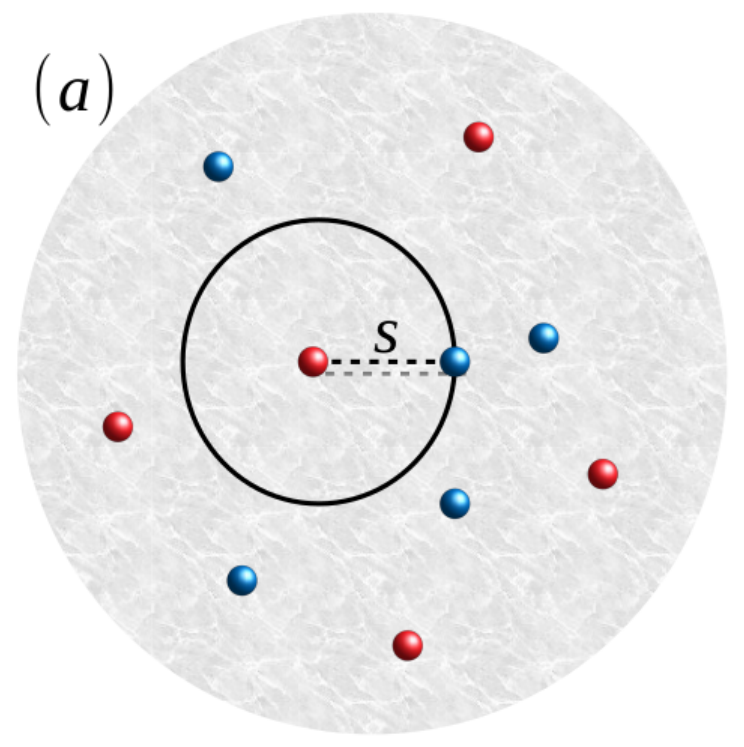}
\includegraphics[width=0.38\columnwidth,angle =0]{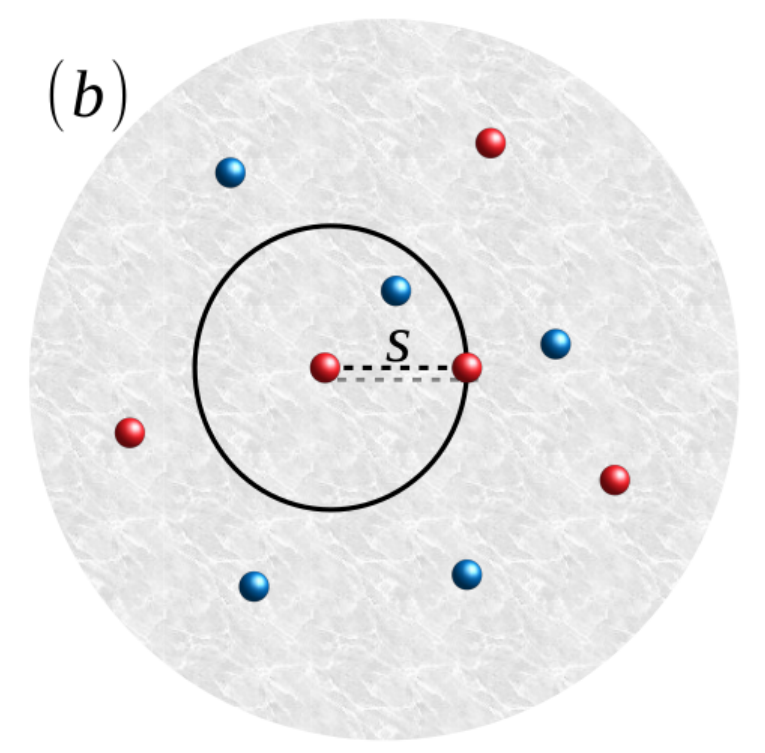}
       \caption{Schematic representation of the spacing $s$ between a reference vortex, at the center of the black circle, and its nearest neighbor with a given topological charge. 
       Red and blue points represent positively and negatively charged vortices, respectively. The associated probability distributions (a) $P_{+-}(s)=P^{(1)}(s)$ and (b) $P_{++}(s)\approx P^{(2)}(s)$ involve conditioning on opposite and equal topological charges of the vortex pair.  
       }
    \label{fig:cartoon2}
\end{figure}

We denote by $P_{ww'}(S)dS$ the probability that given a $w$-vortex of reference, the nearest  $w'$-vortex is at a distance between $S$ and $S+dS$, with no condition on the location of the other $w$-vortices, and with all other $w'$-vortices being further away. See Fig.~\ref{fig:cartoon2} for a schematic representation. 
The specific sign of the charge should have no bearing, so we expect  $P_{++}(S)=P_{--}(S)$, and similarly, $P_{+-}(S)=P_{-+}(S)$.

To estimate $P_{ww'}(S)$, we combine energetic considerations with the PPP-KZM. 
In the case of $P_{++}(S)=P_{--}(S)$, there is likely to be a $-w$ vortex on the interval $(0,S)$ due to the attractive interaction of two oppositely charged vortices. This is verified in Fig.~\ref{fig:densitytq90}. We analytically estimate the probability distribution in this case in Appendix~\ref{app:ppmm}. The distribution is approximately given by $P^{(2)}(S)$, i.e., 
\begin{eqnarray}
P_{++}(S)=P_{--}(S)=2r^4 S^3\exp\left(-r^2 S^2\right),
\label{eq:poissonianppmm}
\end{eqnarray}
where $r=\Gamma (\frac{5}{2})=\frac{3\sqrt{\pi}}{4}$. This indicates that the universal KZM scaling
\begin{eqnarray}
\langle s\rangle_{++}=\langle s\rangle_{--}=\frac{3\sqrt{2\pi}}{4} \hat{\xi}\propto \left(\frac{\tau_{Q}}{\tau_0}\right)^{\frac{ \nu}{1+z\nu}}
\label{eq:meansppmm}
\end{eqnarray}
still holds. 

In the case of $P_{+-}(S)$, it is unlikely to find a vortex with $w=+1$  on the interval $(0,S)$ due to the repulsive interaction between the identically charged vortices. A similar argument applies to $P_{-+}(S)$. Hence, the distribution function obeys $P_{+-}(S)=P_{-+}=P(S)$ and the mean spacing is given by $\langle s\rangle_{+-}=\langle s\rangle_{-+}=\sqrt{2}\langle s\rangle$, where $P(S)$ takes the Wigner-Dyson form in Eq.~\eqref{eq:poissonian}. 
The power-law scaling of the mean spacing with and without conditioning on the winding number is shown in Fig.  \ref{fig:defectscumulants}. Numerical fits to a power law agree with the prediction in Eq. (\ref{eq:meansppmm}). Fitted power-law exponents are consistent with the predicted value $\frac{ \nu}{1+z\nu}=1/4$ within the error bars.

\begin{figure}
 \includegraphics[width=1\columnwidth,angle =0]{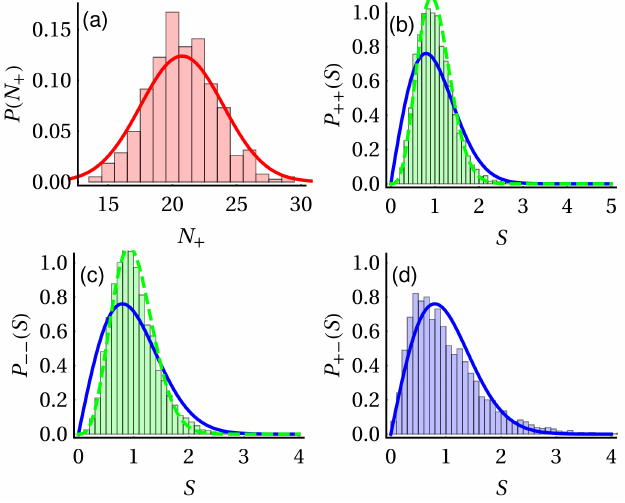}
       \caption{Histogram of the (a) total number of positively charged vortices $N_{+}$, (b) the spacing between the defects  $P_{++}(s)$, (c) $P_{--}(s)$, and (d) $P_{+-}(s)$  at $t=t_{\text{eq}}$ for $\tau_Q=90$.
       The solid red line in (a) represents the distribution Eq.~\eqref{eq:binomialLN}, the dashed green line in (b)-(c) denotes the Eq.~\eqref{eq:poissonianppmm}, and in (b)-(d) the solid blue line corresponds to the distribution \eqref{eq:poissonian}. 
       }
    \label{fig:pdf_ndpm_teq}
\end{figure}

Numerically, one can further identify the charges of each vortex and measure the probability distributions $P_{++}(S)$, $P_{--}(S)$, $P_{+-}(S)$ and $P_{-+}(S)$. Figure~\ref{fig:pdf_ndpm_teq} shows the histogram of positively charged topological defects along with the probability distributions $P_{++}(S)$, $P_{--}(S)$ and $P_{+-}(S)$. The probability distribution $P_{-+}(S)$ is not shown as it exhibits a similar behavior to $P_{+-}(S)$. The distribution $P_{++}(S)=P_{--}(S)$ agrees well with the analytically estimated distributions Eq.~\eqref{eq:poissonianppmm}. The small 
deviation of $P_{+-}(S)$ from the analytical estimate in Eq.~\eqref{eq:poissonian} is likely due to the small but finite probability of finding two equally charged vortices nearby. 
Similarly, the deviation in $P(N_{+})$ is a due to finite size effects, justified in Fig.~\ref{fig:confidencebandcharge}(a), associated with the small number of positively charged vortices. We have verified that similar distributions are observed at different final times $t\sim t_{\text{eq}}$. 

Fig.~\ref{fig:defectscumulants} shows the 
mean spacing $\langle s \rangle_{++}$ and $\langle s \rangle_{+-}$ calculated at $t=t_{\text{ eq}}$ as a function of $\tau_Q$. In the slow quench regime, each of the average spacings  exhibits the universal scaling  with $\tau_{Q}$,  $\langle s \rangle_{++}\propto \langle s \rangle_{+-}\propto \langle s \rangle \propto \tau_{Q}^{1/4}$, in agreement with Eqs. (\ref{eq:means}) and (\ref{eq:meansppmm}).

\subsubsection{Voroni Cell Area Statistics}

   \begin{figure}
 \includegraphics[width=0.99\columnwidth,angle =0]{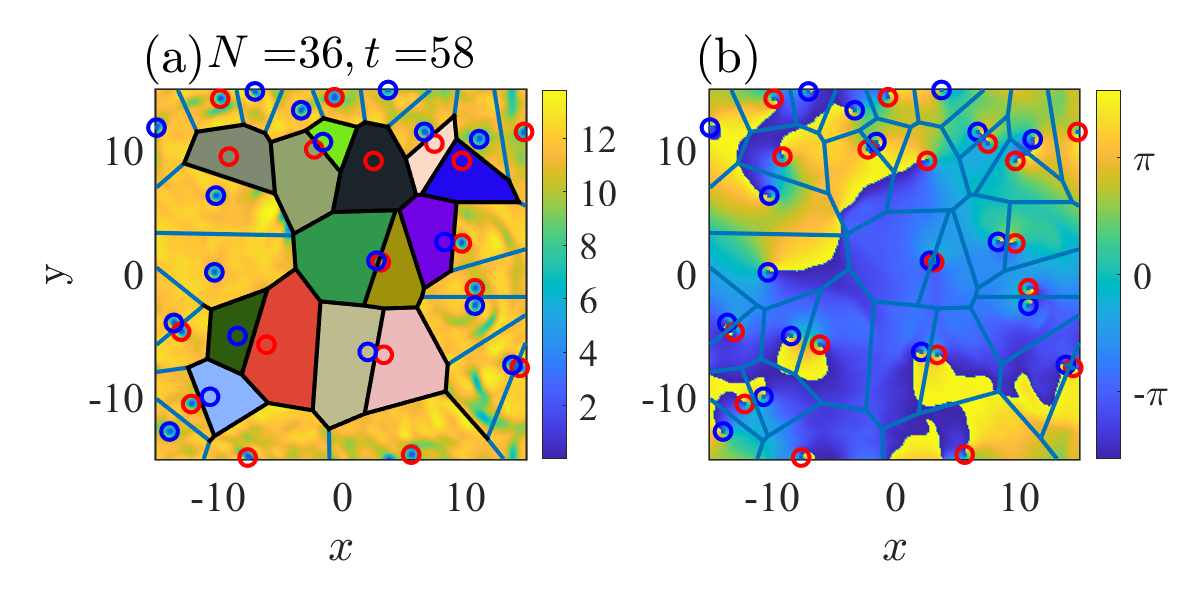}\\
  \includegraphics[width=0.49\columnwidth,height=0.35\columnwidth, angle =0]{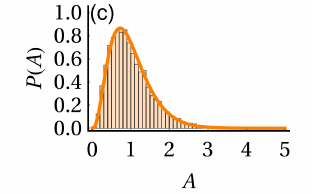}
  \hspace{-1cm}
 \includegraphics[width=0.49\columnwidth,angle =0]{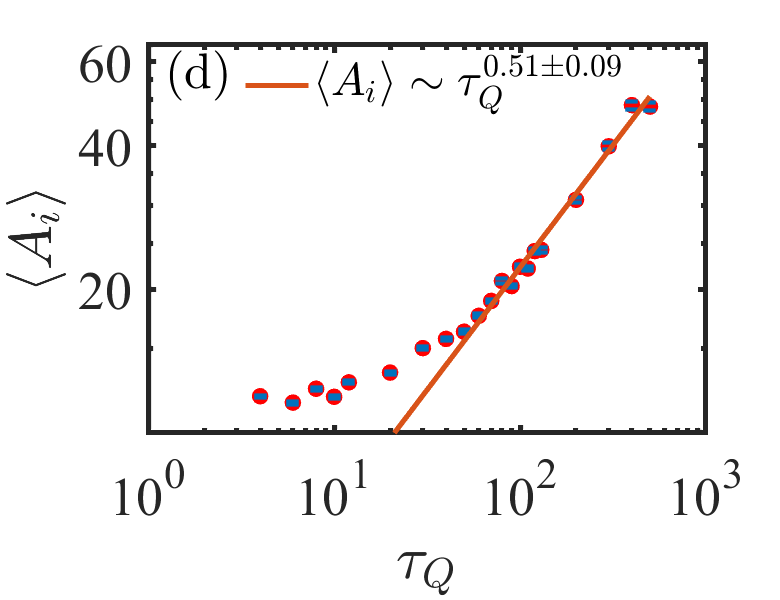}
       \caption{Universal stochastic geometry of a newborn BEC. (a) Voronoi tesselation of a snapshot of the nonequilibrium condensate density $|\Psi|^2$, displaying the spontaneously formed vortices. (b) Voronoi diagram with the corresponding phase $\theta(\mathbf{r},t)$ of the order parameter.   (c) Histogram of the Voronoi cell area distribution at $t = t_{\text{eq}}$ for $\tau_Q=90$, in close agreement with the predicted distribution, Eq.~\eqref{eq:voronoidis}, shown as a solid orange line. (d) The mean Voronoi cell area $\langle A_i \rangle$, depicted in red symbols with error bars, as a function of $\tau_Q$ and evaluated at $t=t_{\text{eq}}$. The solid red line represents the fit $\langle A_i \rangle \propto\tau_Q^{0.51 \pm 0.09 }$, which is consistent with the prediction in Eq. (\ref{eq:voronoimean}), where $\nu = 1/2$ and $z = 2$. The estimation of the Voronoi cell area excludes the cells with vertices beyond the domain of the condensate $[-L,L]\times[-L,L]$, where $L=15$, and use an averaged over $\mathcal{R}=400$ different runs. 
       }
    \label{fig:densityvoronoitq90}
\end{figure}

Voronoi diagrams are an essential tool in stochastic geometry to characterize spatial patterns of point processes \cite{Chiu2013stochastic}. They are helpful in the visualization and characterization of topological defects \cite{Nelson02book}. In particular, they have been used to analyze the spatial statistics of vortices in two-dimensional classical fluids, see, e.g.,   \cite{ReichhardtReichhardt03,Jimenez21}, and the KZM \cite{Stoop18,Reichhardt22,Reichhardt23}.
 
Consider a BEC extended over a domain $\mathcal{D}$ (the support of the density profile) and proliferated with spontaneously formed vortices at the coordinates $\{\mathbf{r}_i\}$. Given the Euclidean distance $d(\mathbf{r},\mathbf{r}_i)=\|\mathbf{r}-\mathbf{r}_i\|$, a Voronoi cell $R_i$ is associated with the vortex at location $\mathbf{r}_i$, by considering the set of points $\mathbf{r}$ that are closer to $\mathbf{r}_i$ than to any other vortex. 
Said differently, 
\begin{equation}
 R_i=\{\mathbf{r}\in \mathcal{D}\,|\,d(\mathbf{r},\mathbf{r}_i)\leq d(\mathbf{r},\mathbf{r}_j)\, \forall j\neq i\}.   
\end{equation}
The union of the Voronoi cells provides a partition of the BEC that tesselates the domain $\mathcal{D}$.
Each Voronoi cell has an area $A_i$. We focus on the cell area distribution $P(A)$, where $A=A_i/\langle A_i \rangle$, and $\langle A_i \rangle$ represents the mean cell area. 
The Voronoi cell area distribution  of a PPP on the plane is approximately given by the  gamma distribution \cite{Weaire:1986Weaire,FERENC2007518}
\begin{eqnarray}
     P(A)=\frac{b^a}{\Gamma(a)} A^{a-1}\exp\left(-b A\right),
 \label{eq:voronoidis}
\end{eqnarray}
where among the different known parameterizations, $a$ and $b$ can be approximately chosen  to be 3.6. This estimation is based on the fact that the average number of edges of the Voronoi cells is 6, and each hexagon cell has the area $A_i = \frac{\sqrt{3}}{2}(s/6)^2$, where $s$ represents the nearest neighbor distance \cite{Weaire:1986Weaire}. 

The mean of Voronoi cell area for a hexagon is $\langle A_i \rangle = \frac{\sqrt{3}}{2} \langle \big(\frac{s}{6}\big)^2 \rangle$. This leads to the universal power-law scaling of the mean Voronoi cell area in a newborn BEC with the quench time, 
\begin{eqnarray}
    \langle A_i \rangle =\frac{\sqrt{3}}{72} \hat{\xi}^2 \propto \left(\frac{\tau_{Q}}{\tau_0}\right)^{\frac{ 2\nu}{1+z\nu}}.
 \label{eq:voronoimean}
\end{eqnarray}

An instance of a Voronoi space tessellation is shown in Fig.~\ref{fig:densityvoronoitq90}(a-b) for a fixed $t = t_{\text{eq}}$ and $\tau_Q=90$. We calculated the average number of edges of the Voronoi cells and found an approximate value 5.7. This value is close to 6, the average number of nearest points predicted by Euler's theorem. The area cell distribution agrees well with the gamma distribution Eq.~\eqref{eq:voronoidis} as shown in Fig.~\ref{fig:densityvoronoitq90}(c) and found $a=b\approx 3.6$. We have further verified in the appendix \ref{app:pppvoronoi}, that the same gamma distribution is obtained by numerically sampling a PPP. In addition, Fig.~\ref{fig:densityvoronoitq90}(d) validates the beyond-KZM universality  $\langle A_i \rangle \propto \tau_Q^{\frac{2\nu}{1+z\nu}}$. 

In short, we have verified the universality of the stochastic geometry characterizing spontaneous vortex patterns in a nonequilibrium BEC formed in finite time. Specifically, we have shown that the predictions of the PPP-KZM model accurately account for the spatial statistics in numerical simulations of the SGPE in a homogeneous BEC, in the absence of an external trap and with periodic boundary conditions. 


\section{Universal vortex statistics for a trapped condensate}
\begin{figure}
  \includegraphics[width=0.96\columnwidth,angle =0]{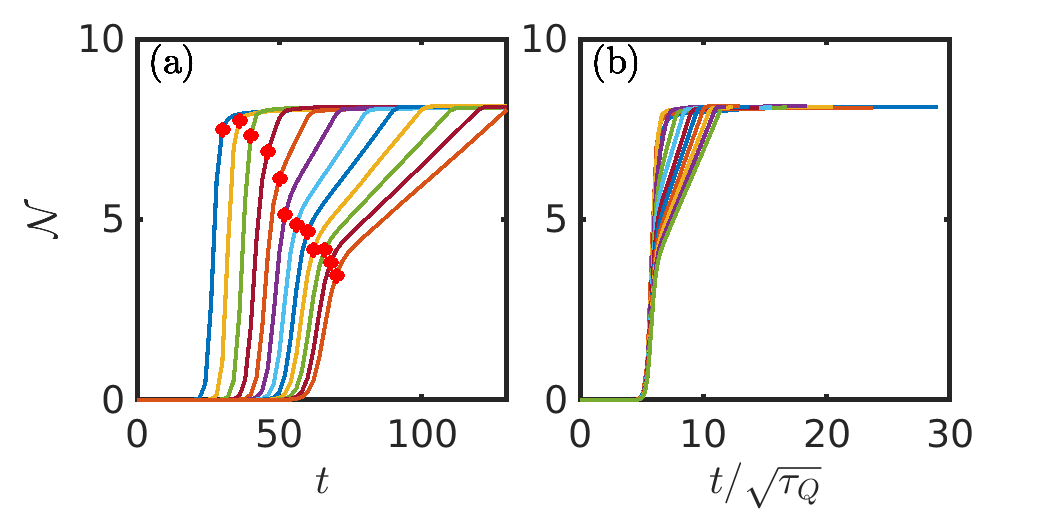}
       \caption{Kibble-Zurek universality in the response time of the order parameter in a BEC transition for a trapped condensate. a) Norm $\mathcal{N}$ as a function of time for different $\tau_Q=(20,...,130)$ (left to right) and $\mathcal{R}=1$, where red points indicate the crossover from an exponential to linear growth and the corresponding time is $t_{\text{eq}}$, (b) Collapse of the growth of the BEC quantified by $\mathcal{N}$ as a function of time scale in terms of the freeze-out time $\hat{t}\propto\sqrt{\tau_Q}$.  For slow quenches ($\tau_Q > 20$), all the lines collapse to a single line for $t<t_{\text{eq}}$. 
       }
    \label{fig:tqalltrap}
\end{figure}

We next show the robustness of the findings reported in the preceding sections in the presence of an uniform  trap. 
To this end, 
we report numerical experiments for a condensate formed in the presence of an external potential.
In such a setting, additional annihilation mechanisms are possible, as vortices may disappear at the edges of the BEC cloud. To validate our results, we consider a hard-wall potential of the form 
\begin{equation}
V(\mathbf{r})= V_0 \Theta\left(\frac{r^2}{a^2}-1\right),
\end{equation}
where $r=\|\mathbf{r}\|$, $\Theta(x)$ is the Heaviside function, and $V_0$ is the trap strength. A similar potential is used in many cold-atom experiments 
\cite{Gaunt13,Navon15,Chomaz15,Navon16,Mukherjee17,Guillaume2019giant}. For our numerical simulations, we fix $V_0=60 \gg \mu$.
As in the case of the homogeneous BEC,  we 
begin our numerical experiments with the initial condition $\Psi=0$ and let the system relax 
for a time $t_{0}=10$ before the beginning of a quench. Figure~\ref{fig:tqalltrap}(a) shows the response time of the order parameter in a BEC transition. The collapse shown in Fig.~\ref{fig:tqalltrap}(b) upon rescaling the evolution time by $\sqrt{\tau_Q}$ confirms the universal scaling law predicted by the KZM for the freeze-out time  $\hat{t}\propto\sqrt{\tau_Q}$ with mean-field critical exponents $\nu = 1/2$ and $z = 2$.  While a steepening of the power-law scaling of the defect number with the quench time is expected in the presence of an extended external confining potential \cite{DRP11,DKZ13}, the validity of the standard KZM scaling indicates that the transition remains effectively homogeneous in the presence of the box-like trap, i.e., unaffected by the Inhomogeneous KZM \cite{DKZ13}.

    \begin{figure}
 \includegraphics[width=1.0\columnwidth,angle =0]{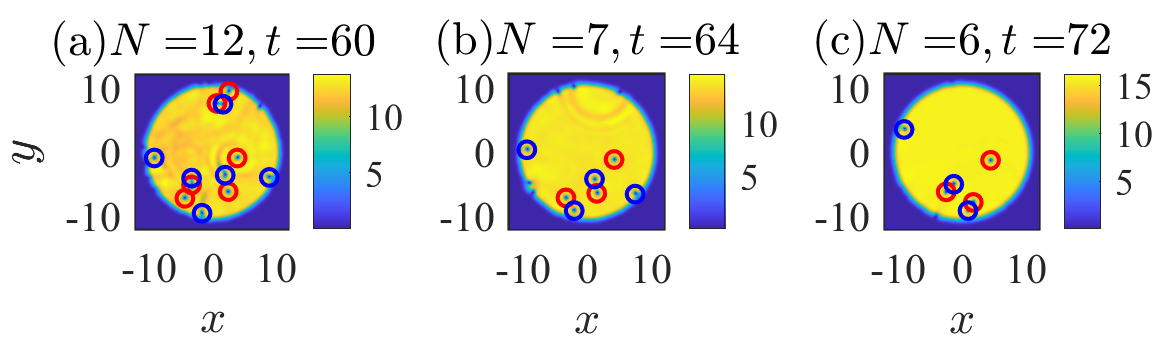}
       \caption{Snapshots of the nonequilibrium condensate density $|\Psi|^2$ showing the spontaneously formed vortices in a homogeneous 2D BEC in a trap at different times ($t \ge t_{\text{eq}}$) for $\tau_Q=90$. Red and blue symbols represent positive and negatively charged vortices, respectively,  with $N$ denoting the total vortex number. 
       }
    \label{fig:densitytq90trap}
\end{figure}
    \begin{figure}
  \includegraphics[width=0.98\columnwidth,angle =0]{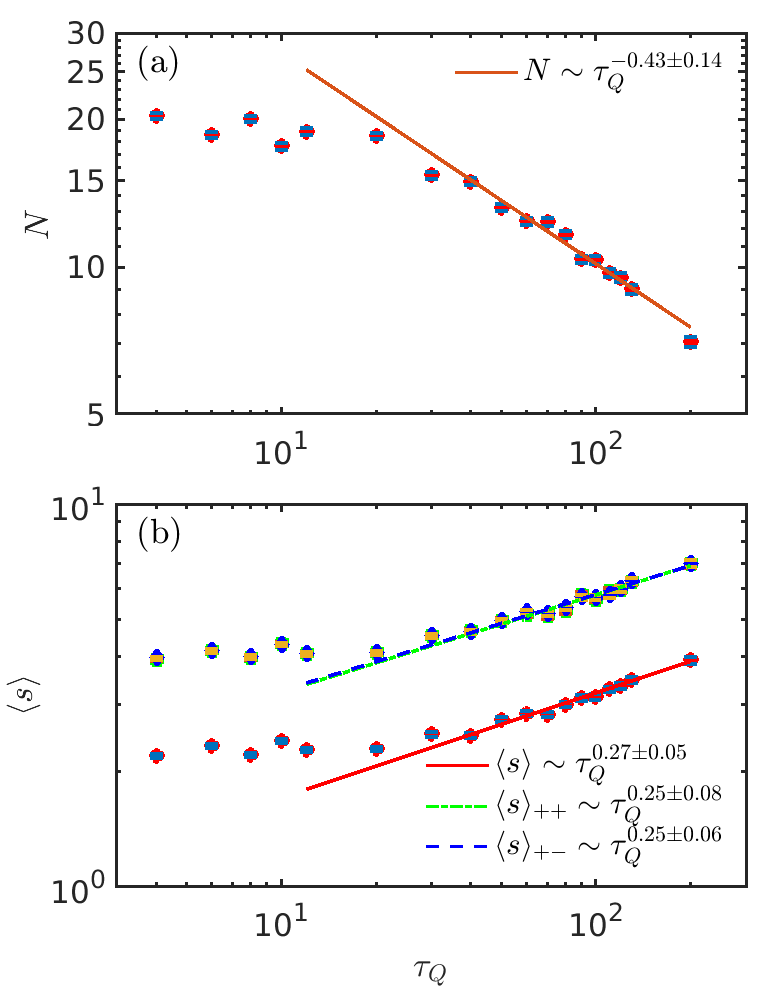}
       \caption{BEC formation in a hard-wall trap. (a) The total number of defects $N$ and (b)  the mean spacing $\langle s \rangle$, $\langle s \rangle_{++}$ and $\langle s \rangle_{+-}$  in the trap as a function of $\tau_Q$ produced at $t=t_{\text{eq}}$ averaged over $\mathcal{R}=200$ different runs are shown in red symbols with error bars. }
    \label{fig:defectscumulantstrap}
\end{figure}
Figure~\ref{fig:densitytq90trap} shows different snapshots of the condensate density $|\Psi|^2$ during the Bose-Einstein condensation. As in the homogeneous case, the newborn BEC is proliferated by positively and negatively charged vortices formed at different times of the dynamical process. Unlike in the homogeneous condensate with periodic-boundary conditions, the total vortex charge in the presence of the box-like trap need not vanish due to the decay of vortices through the periphery. Nevertheless, due to the hard-wall trap,  this process is minimized by the mesa-like profile of the newborn BEC, and the primary mechanism of the vortex annihilation is through antivortex interaction. 
 Fig.~\ref{fig:defectscumulantstrap} shows the average number of the total defects $N$ and mean spacing $\langle s \rangle$ calculated at $t=t_{\text{eq}}$ as a function of $\tau_Q$. In the slow quench regime, the average number of defects exhibits a scaling $N\propto \tau_{Q}^{-0.43\pm 0.14}$, while the mean spacing scales with $\tau_{Q}$ as $\langle s \rangle \propto \tau_{Q}^{ 0.27 \pm 0.05 }$. The scaling law of $N$ is consistent with the KZM prediction of the defect density. On the other hand, both these scaling laws are in good agreement with the predictions in Eqs.~\eqref{eq:kqFCS} (for $q=1$) and ~\eqref{eq:means} beyond the KZM. Moreover, the mean spacing $\langle s \rangle_{++}$
and $\langle s \rangle_{+-}$, Eq.~\eqref{eq:meansppmm}, are also consistent with the KZM predictions. Additional simulations reported in the App. \ref{app:trap} confirm the universal stochastic geometry of vortex patterns in a trapped newborn BEC.

\begin{table*}[!htb]
\begin{tabular}{ |p{5cm}|p{5.5cm}|p{7cm}|  }
\hline
\multicolumn{3}{|c|}{{ \bf Vortex Spatial Statistics and Stochastic Geometry}} \\
\hline
 Name & Mean  & Probability Distribution \\
& & \\
Vortex Distance Distribution & $\langle s \rangle = \frac{128}{45\pi}R$ & $ P_2(s)=\frac{4s}{\pi R^2}\bigg[\arccos\left(\frac{s}{2R}\right)-\frac{s}{2 R} \sqrt{1-\frac{s^2}{4R^2}}\bigg]$ \\
& & \\
Vortex Spacing distribution & $\langle s\rangle=\frac{\sqrt{\pi}}{2}\hat{\xi}\propto \left(\frac{\tau_{Q}}{\tau_0}\right)^{\frac{ \nu}{1+z\nu}}$ & $P(S)=P^1(S)=\frac{\pi}{2}S\exp\left(-\frac{\pi}{4}S^2\right)$ \\
& & \\
Vortex Spacing Distribution of $k$-th Order & $\langle s\rangle=\frac{\Gamma{(\frac{1}{2}+k)}}{\Gamma{(k)}}\hat{\xi}\propto \left(\frac{\tau_{Q}}{\tau_0}\right)^{\frac{ \nu}{1+z\nu}}$   & $P^{(k)}(S)=  \frac{2}{(k-1)\,!} r^{2k-1}S^{2k-1}e^{- r^2 S^2},~~r=\frac{\Gamma \big(\frac{1}{2}+k \big)}{\Gamma (k)}$ \\
& & \\
Conditioned Vortex-Antivortex Spacing Distributions & 
$\langle s\rangle_{+-}=\langle s\rangle_{-+}
=\sqrt{2\pi}\hat{\xi}\propto \left(\frac{\tau_{Q}}{\tau_0}\right)^{\frac{ \nu}{1+z\nu}}$
& 
$P_{+-}(S)=P_{-+}(S)=P(S)$
\\ 
& & \\
Conditioned Vortex-Vortex Spacing Distributions & 
$\langle s\rangle_{++}=\langle s\rangle_{--}=\frac{3\sqrt{2\pi}}{4} \hat{\xi}\propto \left(\frac{\tau_{Q}}{\tau_0}\right)^{\frac{ \nu}{1+z\nu}}$
& 
$P_{++}(S)=P_{--}(S)=2r^4 S^3\exp\left(-r^2 S^2\right),
r=\Gamma (\frac{5}{2})$
\\ 
& & \\
Voronoi Cell Area Statistics    &  $\langle A_i \rangle =\frac{\sqrt{3}}{72} \hat{\xi}^2 \propto \left(\frac{\tau_{Q}}{\tau_0}\right)^{\frac{ 2\nu}{1+z\nu}}$ & $P(A)=\frac{b^a}{\Gamma(a)} A^{a-1}\exp\left(-b A\right), ~~A=A_i/\langle A_i\rangle$\\ & & \\
\hline
\end{tabular}
\caption{Summary of the results for the vortex spatial statistics and stochastic geometry.}
\label{Table:vortexspacing}
\end{table*}

\section{Conclusions}
A continuous phase transition crossed in finite time leads to the formation of topological defects. Decades of research have established the validity of the KZM in predicting the mean density of defects, but are there universal features of the critical dynamics beyond its scope?
A positive answer to this question has been the subject of a recent series of studies establishing the universality of the defect number distribution under the umbrella of beyond-KZM physics in both quantum \cite{delcampo18,Cui20,Bando20,King22,Gherardini23} and classical systems \cite{GomezRuiz20,delcampo21,Mayo21,Subires22,GomezRuiz22}.

In this work, we have addressed a complementary aspect regarding the spatial statistics of spontaneously formed topological defects. In particular, we have focused on the vortex formation during Bose-Einstein condensation of a Bose gas. 
In this context, the average number of defects is well described as a function of the cooling rate by KZM in the limit of slow cooling. Vortex number fluctuations are shown to be universal. The vortex number distribution is well described by a binomial distribution with a mean predicted by the KZM. Specifically, numerical simulations of a homogeneous and trapped condensate lead to a 
normal vortex number distribution. The dynamics of Bose-Einstein condensation results in vortex-antivortex pairs, and the winding number of each vortex is restricted to $w=\pm1$.

The vortex spatial statistics can be described as a PPP on a plane with KZM density. The vortex distance distribution normalized to the mean are equivalent to those in the celebrated Disk Line Picking problem and admit closed-form analytical expressions in excellent agreement with SGPE simulations. The vortex spacing distribution is described by a Wigner-Dyson distribution, when vortices and antivortices are treated on equal footing. However, when the spacing distribution is conditioned on the topological charge, it changes from Wigner-Dyson to  
that of the next nearest neighbor spacing distribution, i.e., the spacing distribution of second order. The agreement with the PPP-KZM model extends beyond two-point correlations and is established for the $k$-th nearest neighbor distribution. 
Further evidence of the universal vortex stochastic geometry is provided by considering the random tessellation of the spontaneous vortex distribution. The corresponding Voronoi area cell statistics follows a universal gamma distribution, with universal mean area cell scaling with the quench rate. 

Our results, summarized in Table \ref{Table:vortexspacing}, establish the universality of the emergent stochastic geometry of spontaneously formed point-like topological defects generated across a continuous phase transition. In doing so, they unveil universal signatures of the critical dynamics of systems driven across a phase transition that lie beyond the scope of the KZM.  These predictions are  testable in a wide variety of experiments involving BEC formation, as well as in other scenarios characterized by the spontaneous creation of point-like topological defects, such as kinks and vortices, e.g., in superfluids, superconductors, quantum fluids of light \cite{Carusotto13}, colloidal systems \cite{Keim15}, and multiferroics \cite{Lin14}.  Applications exploiting these findings can be envisioned, e.g., in the study of turbulence, structure formation, tribology, and the engineering of functional materials.

 \section*{Acknowledgments}
 The authors are indebted to Seong-Ho Shinn for insightful comments on the manuscript. 
 It is a pleasure to acknowledge discussions with Fernando G\'omez-Ruiz, Masahito Ueda,  and Hai-Qing Zhang. This research was funded in part by the Luxembourg National Research Fund (FNR), grant reference [17132060]. For the purpose of open access, the author has applied a Creative Commons Attribution 4.0 International (CC BY 4.0) license to any Author Accepted Manuscript version arising from this submission.
\appendix
\section{Probability Distribution $P^{(k)}(s)$ for $k$-th Nearest Neighbor} \label{app:ppmmkth}
In this appendix, we compute the probability $P^{(k)}(s)$ of finding a $k$-th nearest neighboring vortex at a distance $s$ from a reference vortex. It is known that for a Poisson point process on a disk of radius $R$, the $k$-th nearest neighbor spacing distribution reads \cite{mathai1999introduction} 
\begin{eqnarray}
P^{(k)}(s)ds=2\frac{(\lambda \pi)^k}{(k-1)\,!} s^{2k-1}e^{-\lambda \pi s^2}ds,~~0\le s \le R,
\label{eq:kneighbor}
\end{eqnarray}
where the $\lambda$ is the intensity, {i.e}, the number of  points in an area $A$.

 For $N$ random vortices on a disk of radius $R$ with area $A=\pi R^2$, the intensity $\lambda=\frac{N}{\pi R^2}$. Thus, ~\eqref{eq:kneighbor} becomes
\begin{eqnarray}
P^{(k)}(s)ds=2 N^k R^{-2k} \frac{1}{(k-1)\,!} s^{2k-1}e^{-\frac{N}{R^2}  s^2}ds,
\end{eqnarray}
for $0\le s \le R$.
On rescaling $s=xR$, we get 
\begin{eqnarray}
P^{(k)}(s)ds=  \frac{2 N^k}{(k-1)\,!} x^{2k-1}e^{- N x^2}dx,~~0 \le x \le 1.
\end{eqnarray}
The average value of the $k$-th order spacing distribution is thus given by
\begin{eqnarray}
\langle s \rangle &=&\int_0^{R} sP^{(k)}(s)ds\\&=& N^{-1/2}R \frac{\Gamma \big(\frac{1}{2}+k \big)}{\Gamma (k)}\nonumber\\&=&\hat{\xi} \frac{\Gamma \big(\frac{1}{2}+k \big)}{\Gamma (k)},\nonumber
\end{eqnarray}
where $R=N^{1/2} \hat{\xi}$. For $S=s/\langle s \rangle = xR/\langle s \rangle$,
\begin{eqnarray}
P^{(k)}(S)dS=  \frac{2}{(k-1)\,!} r^{2k}S^{2k-1}e^{- r^2 S^2}dS,
\label{eq:kneighborf}
\end{eqnarray}
where $r=\frac{\Gamma \big(\frac{1}{2}+k \big)}{\Gamma (k)}$. 

For $k=1$, this $k$-th neatest neighbor spacing distribution reduces to the prediction put forward in Ref. \cite{delcampo22} and takes the form of the well-known Wigner-Dyson distribution Eq.~\eqref{eq:poissonian}.

For a $d$-ball, a systematic derivation can be done from the probability distribution
 \begin{equation}
   P^{(k)}(s)ds=Q(s)R(s) g(s)ds,
\end{equation}
 where 
 \begin{eqnarray}
g(s)&=& \binom{N-k}{1} \frac{S_{d-1}(s)}{V_d(R)},\\
R(s)&=&\binom{N-1}{k-1} \bigg(\frac{V_d(s)}{V_d(R)}\bigg)^{k-1},\\
Q(s)&=&\bigg(1-\frac{V_d(s)}{V_d(R)}\bigg)^{N-k-1},
\end{eqnarray}
with the volume $V_d(x)$ and surface $S_{d-1}(x)$ being  defined as
 \begin{equation}V_d(x)=\frac{\pi^{d/2}}{\Gamma (\frac{d}{2}+1)}x^d,~~S_{d-1}(x)=\frac{2\pi^{d/2}}{\Gamma (\frac{d}{2})}x^{d-1},\end{equation}
 in terms of the Gamma function $\Gamma (x)$.
 
Thus, $P^{(k)}(s)$ has different contributions.
Out of $N$ vortices, one is taken as a referece, $k-1$ out of the remaining $N-1$ are chosen as the first neighbors in the interval $(0,s)$ with probability $R(s)$, the $k$-th neighbor is chosen out of the remaining $N-k$ vortices at a spacing between $s$ and $s+ds$ with probability $g(s)$, with all other vortices are found further away with probability $Q(s)$. 
For large $N$, this leads to the equation 
 \begin{eqnarray}
P^{(k)}(S)dS=  \frac{d}{(k-1)\,!} r^{dk}S^{dk-1}e^{- r^d S^d}dS,
\label{eq:kneighborf}
\end{eqnarray}
where $r=\frac{\Gamma \big(\frac{1}{d}+k \big)}{\Gamma (k)}$.
\section{Distribution of Poisson Point Process}\label{app:ppp}

\begin{figure}[t]
 \includegraphics[width=0.98\columnwidth,angle =0]{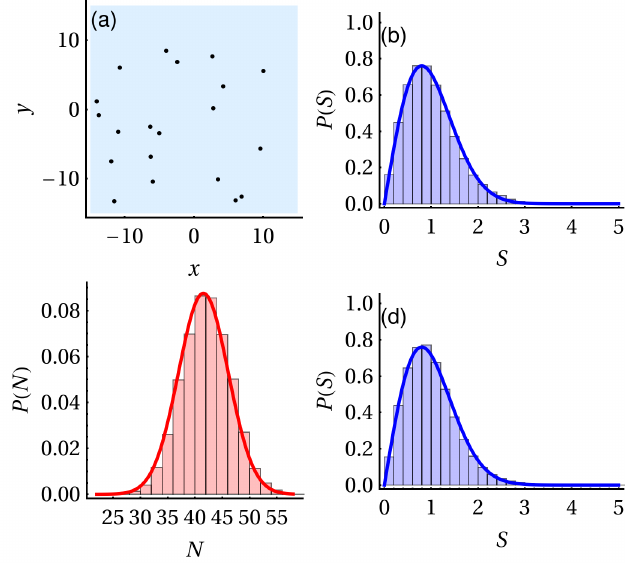}
       \caption{(a) Random points on a domain $[-15,15]\times[-15,15]$, (b) histogram of the spacing between the defects for a fixed $N$ total number of random points, (c) histogram of the total number of random points $N$ obtained from a Poisson point process for $\mathcal{R}=10000$ different initial conditions and (d) the corresponding histogram of the spacing between the defects, where the solid blue line in (b) and (d) represents the distribution Eq.~\eqref{eq:poissonian}. The solid red line in (c) denotes the distribution Eq.~\eqref{eq:binomialLN}. 
       }
    \label{fig:pdf_ppp}
\end{figure}

This appendix compares the probability distribution obtained from a PPP for a fixed $N$ and varying $N$. Fig.~\ref{fig:pdf_ppp} shows the corresponding histogram obtained from  $\mathcal{R}=10000$ different initial conditions for fixed and varying total number $N$ of random points generated on a domain of size $[-15, 15] \times [-15, 15]$. 
The difference in the spatial statistics with fixed and varying $N$ is minimal due to the interdependent Poisson process for each $N$. Moreover, the probability distribution $P(s)$ matches well with the Wigner-Dyson distribution, Eq.~\eqref{eq:poissonian}.

\section{Confidence Bands and Estimation for Probability Distribution of Spatial Statistics}\label{app:confidenceband}
    \begin{figure}[h]
 \includegraphics[width=0.98\columnwidth,angle =0]{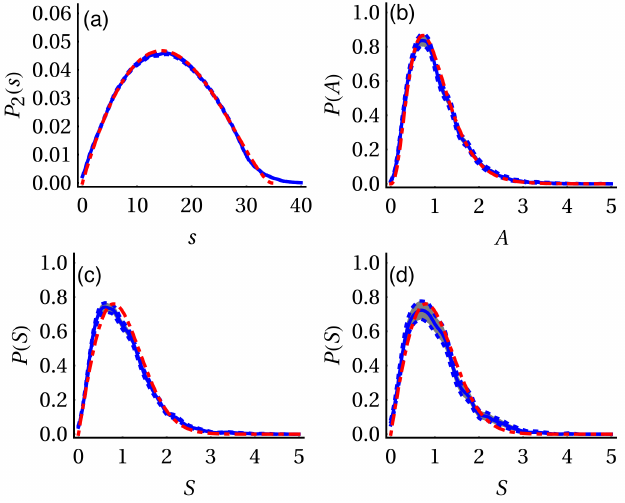}
       \caption{Probability density function of a) distance $s$ between two defects,  (b) the Voronoi cell area distribution, (c) spacing between topological defects $S$ for varying $N$, and (d) and the same for fixed N at $t = t_{\text{eq}}$ for $\tau_Q=90$ of a homogeneous condensate. The thick blue solid line represents the numerical estimates, and the red dot-dashed line denotes the corresponding analytical estimates. The dashed blue lines represent the confidence interval. 
       }
    \label{fig:confidenceband}
\end{figure}
    \begin{figure}[h]
 \includegraphics[width=0.98\columnwidth,angle =0]{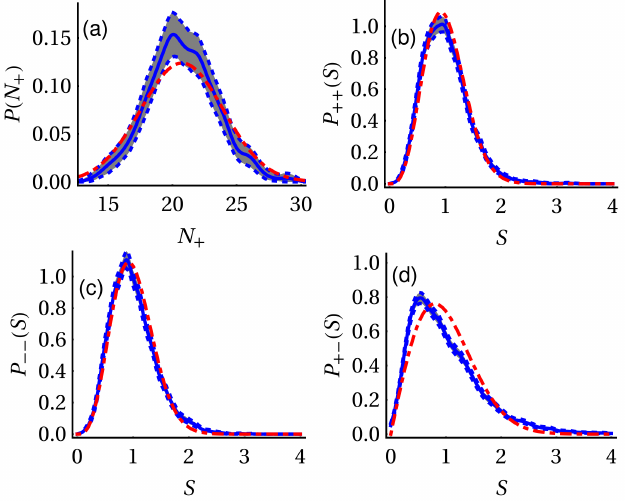}
       \caption{Histogram of the (a) total number of positively charged vortices $N_{+}$, (b) the spacing between the defects  $P_{++}(s)$, (c) $P_{--}(s)$, and (d) $P_{+-}(s)$  at $t=t_{\text{eq}}$ for $\tau_Q=90$. The dot-dashed red line in (a) represents the distribution Eq.~\eqref{eq:binomialLN}, the same in (b)-(c) denotes the Eq.~\eqref{eq:poissonianppmm}, and in (b)-(d) corresponds to the distribution \eqref{eq:poissonian}. The thick blue solid line represents the numerical estimates, and the dashed blue lines represent the confidence interval.
       }
    \label{fig:confidencebandcharge}
\end{figure}

Fig.~\ref{fig:confidenceband} shows the probability distributions of (a) the distance $s$ between two defects, (b) the Voronoi cell area distribution, (c) the spacing between topological defects $S$ for varying $N$, and (d) and the same for fixed $N$ at $t = t_{\text{eq}}$ for $\tau_Q=90$ of a homogeneous condensate with confidence intervals. The blue solid lines represent the non-parametric density estimate (smoothed kernel) and blue dashed lines denote bootstrap-created confidence bands. The results show that kernel densities match the analytical estimate (dot-dashed red lines) well. 

We additionally compare the analytical and numerical estimates of various probability distributions for conditioned vortices by representing the 
distributions with the inclusion of confidence bands as shown in Fig. \ref{fig:confidencebandcharge}. We see the qualitative matching between numerically calculated kernel densities and the analytical estimate (dot-dashed red lines). 

\section{Probability Distributions $P_{++}$ and $P_{+-}$} \label{app:ppmm}
In this appendix, we compare the total vortex spacing distribution $P(s)$ with the probability distributions $P_{++}(s)$ and $P_{+-}(s)$ that involve conditioning on the winding number (circulation) of the vortices.

Let $g(s)ds$ be the conditional probability of finding a vortex on a disk in the interval $[s,s+ds]$ given that there is a vortex at the origin $s=0$. The maximum radius of the disk $(d=2)$ is $R$. 
The total probability of finding any of the $N-1$ vortices as the nearest neighbor at a distance $s$ from the origin is,
 \begin{equation}
P(s)ds=Q(s)g(s)ds,
\end{equation}
with the normalization condition $\int_0^RP(s)ds=1$,
where $Q(s)$ represents the probability of finding the remaining $N-2$ vortices further away. In this case, the charges of vortices are not considered. Additionally,

 \begin{equation}
g(s)ds= \binom{N-1}{1} \frac{S_{d-1}(s)}{V_d(R)}ds
\end{equation}
and
 \begin{equation}
Q(s)=\bigg(1-\frac{V_d(s)}{V_d(R)}\bigg)^{N-2}.
\end{equation}

\subsection{Probability Distribution $P_{+-}(S)$} 
We next focus on the spacing distribution conditioned on the winding number $w=\pm1$. 
Consider that the reference point is a $w=+1$ vortex. Then, the probability of finding a $w=-1$ vortex becomes 
 \begin{equation}
P_{+-}(s)ds=Q(s)g(s)ds,
\end{equation}
where 
\begin{eqnarray}
g(s)ds&=& \binom{N/2}{1} \frac{S_{d-1}(s)}{V_d(R)}ds,\\
Q(s)&=&\bigg(1-\frac{V_d(s)}{V_d(R)}\bigg)^{\frac{N}{2}-1},
\end{eqnarray}
where we note we do not condition the location of the other vortices with $w=+1$.  
 We next use the rescaled variable
$X=s/R$. We relate $R$ with the correlation length out of equilibrium $\hat{\xi}$ with the expression $R=N^{1/d}\hat{\xi}=N^{1/d} \xi_0   \left(\frac{\tau_{Q}}{\tau_0}\right)^{\frac{ \nu}{1+z\nu}}$ \cite{delcampo22}. In terms of this new variable $X$, $P_{+-}(s)ds$ reads 
 \begin{equation}
P_{+-}(s)ds=d \frac{N}{2}X^{d-1}(1-X^d)^{\frac{N}{2}-1} dX,
\label{eq:psplusminus}
\end{equation}
where $\frac{S_{d-1}(s)}{V_d(R)}=\frac{d}{R}X^{d-1}$ and $\frac{V_d(s)}{V_d(R)}=X^{d}$.   Equation ~\eqref{eq:psplusminus} is normalized, i.e., 
 \begin{equation}
\int_0^{R} P_{+-}(s)ds=\int_0^{1}P(X)dX=1.
\end{equation}

In the large $N$ limit 
 \begin{equation}
P_{+-}(s)ds=d \frac{N}{2} X^{d-1} \exp\left(-\frac{N}{2}X^d\right) dX,
\end{equation}
and the mean spacing 
 \begin{eqnarray}
\langle s \rangle_{+-} &=&\int_0^{R} s P_{+-}(s)ds\nonumber\\
&=&d \hat{\xi} \frac{N^{1+1/d}}{2}\int_0^{1}X^{d} \exp\left(-\frac{N}{2}X^d\right)dX\nonumber\\
&=& 2^{\frac{1}{d}}\hat{\xi}\, \Gamma \left(1 + \frac{1}{d}\right).
\end{eqnarray}
We thus find that the spacing with and without conditioning on the topological charge are related as
\begin{eqnarray}
\langle s \rangle_{+-}=2^{\frac{1}{d}}\langle s \rangle,
\end{eqnarray}
as expected from dimensional analysis.

 We next normalize the distance $s$ with respect to the mean $\langle s \rangle_{+-}$. In the scaled variable $S=s/ \langle s \rangle_{+-} $, 
 \begin{equation}
X = S \langle s \rangle_{+-}/R= r \left(\frac{2}{N}\right)^{\frac{1}{d}}  S,  
\end{equation}
where $r= \Gamma (1 + \frac{1}{d})$. 
As a function of the spacing normalized to the mean, the vortex-antivortex spacing distribution reads
\begin{eqnarray}
P_{+-}(S)dS=d r^{d} S^{d-1} \exp(-r^{d}S^d)dS.
\end{eqnarray}
Hence,
\begin{eqnarray}
P_{+-}(S)=P_{-+}(S)=d r^{d} S^{d-1} \exp(-r^{d}S^d).
\end{eqnarray}
\subsection{Probability Distribution $P_{++}(S)$}
Let us now consider that the reference point is a $w=+1$ vortex. Then, the probability of finding the nearest neighbor  vortex with $w=+1$ becomes 
 \begin{equation}
\label{pppeq} 
P_{++}(s)ds=Q(s)g(s)ds.
\end{equation}
In the PPP-KZM model, all configurations with an arbitrary number of antivortices between the nearest neighbor $w=+1$ vortices should be taken into account and summed over. 
This is equivalent to no conditioning on the location of the $N/2$ antivortices with $w=-1$, resulting in (\ref{pppeq}).
However, such prediction does not describe the numerically simulated data.

In what follows, we deviate from the strict PPP-KZM model and find an approximated estimate of $P_{++}(s)$ that accurately describes the numerics. 
This approximation builds on the fact that the interactions between any pair of defects are attractive or repulsive depending on whether they have opposite or equal sign winding numbers. It is these interactions that are not taken into account in the PPP-KZM model. As a result, two oppositely charged vortices are likely to be closer than two equally charged vortices.

Among all the possible configurations of the $N/2$ antivortices, we find that the numerical data is consistent 
with the configuration in which an antivortex is located between the two nearest neighbor vortices used to define $P_{++}(s)$. Said differently, we focus on the configuration 
with a $w=-1$ defect in $[0,s]$.  Accordingly, we estimate $P_{++}(s)$ as  
 \begin{equation}
P_{++}(s)ds=Q(s)R(s)g(s)ds,
\end{equation}
 where 
 \begin{eqnarray}
R(s)&=&\binom{N/2}{1} \frac{V_d(s)}{V_d(R)},\\
g(s)&=& \binom{N/2-1}{1} \frac{S_{d-1}(s)}{V_d(R)},\\
Q(s)&=&\bigg(1-\frac{V_d(s)}{V_d(R)}\bigg)^{\frac{N}{2}-2}.
\end{eqnarray}

We now use the rescaled variable
$X=s/R=s/(N^{1/d} \hat{\xi})$, where 
$\hat{\xi}=\xi_0   \left(\frac{\tau_{Q}}{\tau_0}\right)^{\frac{ \nu}{1+z\nu}}$.  In this variable 
 $\frac{S_{d-1}(s)}{V_d(R)}=\frac{d}{R}X^{d-1}$ and $\frac{V_d(s)}{V_d(R)}=X^{d}$. This leads to the relation
 \begin{equation}
P_{++}(s)ds=d\frac{N}{2} \left(\frac{N}{2}-1\right)X^{2d-1}(1-X^d)^{\frac{N}{2}-2} dX,
\end{equation}
fulfilling the normalization condition
 \begin{equation}
\int_0^{R} P_{++}(s)ds=\int_0^{1}P(X)dX=1.
\end{equation}

In the large $N$ limit,
 \begin{equation}
P_{++}(s)ds=d\frac{N^2}{4}X^{2d-1} \exp\left(-\frac{N}{2}X^d\right)dX.
\end{equation}
The corresponding mean spacing reads
 \begin{eqnarray}
\langle s \rangle_{++} &=&
\int_0^{R} s P_{++}(s)ds\nonumber\\
&=&d \hat{\xi} N^{2+1/d}\int_0^{1}X^{2d} \exp{(-NX^d)} dX\nonumber\\
&=& 2^{\frac{1}{d}}\hat{\xi}\, \Gamma \left(2 + \frac{1}{d}\right).
\end{eqnarray}
We note that $\langle s \rangle_{++}$ is related to the unconditioned spacing $\langle s \rangle$ as
\begin{equation}
\langle s \rangle_{++}=3\times2^{\frac{1-d}{d}}\langle s \rangle.
\end{equation}
In terms of $S=s/ \langle s \rangle_{++} $, one finds
 \begin{equation}
X = S \langle s \rangle_{++}/R= r \left(\frac{2}{N}\right)^{\frac{1}{d}}   S, 
\end{equation}
where $r=\Gamma (2 + \frac{1}{d})$. 
Using the spacing normalized to the mean, 
\begin{eqnarray}
P_{++}(S)dS=d r^{2d} S^{2d-1} \exp(-r^{d}S^d)dS,
\end{eqnarray}
and thus
\begin{eqnarray}
P_{++}(S)=P_{--}(S)=d r^{2d} S^{2d-1} \exp(-r^{d}S^d).
\end{eqnarray}

\section{Distribution of Voronoi Cell Area of a Poisson Point Process}\label{app:pppvoronoi}
\begin{figure}
\includegraphics[width=0.98\columnwidth,angle =0]{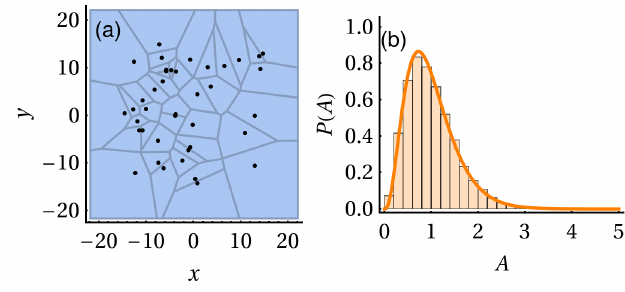}
       \caption{(a) Voronoi diagram of random points on a domain $[-15,15]\times[-15,15]$ and (b) histogram of the Voronoi cell area, where the solid orange line represents the distribution, Eq.~\eqref{eq:voronoidis}.
       }
    \label{fig:pdf_pppvoronoi}
\end{figure}
In this appendix, we compare the probability distribution of Voronoi cell area obtained from a numerically generated PPP  with the analytically estimated probability distribution $P(A)$ given in Eq.~\eqref{eq:voronoidis}. Fig.~\ref{fig:pdf_pppvoronoi} shows the Voronoi diagram of random
points generated on a domain $\mathcal{D}=[-15, 15] \times [-15, 15]$ and the histogram of Voronoi cell area $A_i$. We find an excellent agreement between that theoretically estimated probability distribution, Eq.~\eqref{eq:voronoidis}, and that of a PPP.
\begin{figure*}
 \includegraphics[width=1.95\columnwidth,angle =0]{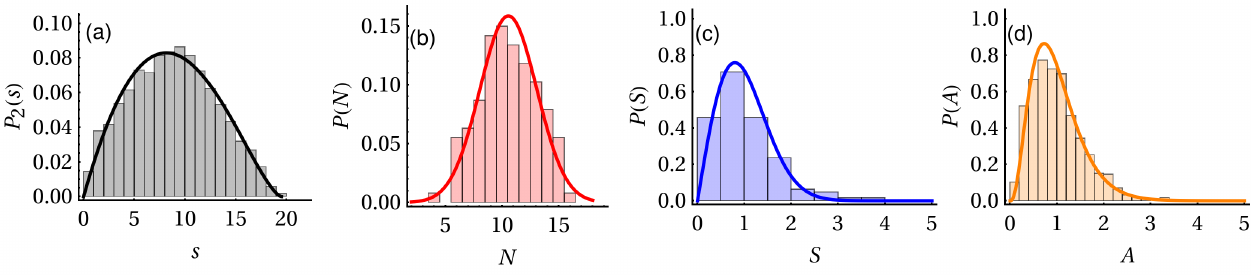}
       \caption{Histogram of the (a) distance $s$ between two defects, (b) total number of defects $N$, (c) the spacing between topological defects $S$ for fixed $N$, and (d) the Voronoi cell area distribution at $t = t_{\text{eq}}$ for $\tau_Q=90$ of a trapped condensate. The solid black line in (a) represents the distribution Eq.~\eqref{eq:dist2}. The solid red, blue, and orange lines in (b), (c), and (d) denote the Eq.~\eqref{eq:binomialLN}, \eqref{eq:poissonian} and \eqref{eq:voronoidis}, respectively.
       }
    \label{fig:pdf_all_trap}
\end{figure*}

\section{Histogram of Defects and Spatial Statistics for a Trapped Condensate}\label{app:trap}

Figure~\ref{fig:pdf_all_trap}(a) shows the histogram of the distance $s$ between two defects at $t=t_{\text{eq}}$ for $\tau_Q=90$. Panels (b) and (c) show the histogram of the total number of defects $N$ and the spacing between topological defects $S$ for a fixed $N$, respectively. As in the case of a homogeneous condensate, the vortex spatial statistics agree well with the respective analytical estimations. Finally, in Fig.~\ref{fig:pdf_all_trap}(d), we show that the histogram of the Voronoi cell area distribution of a trapped condensate also follows the relation Eq. \eqref{eq:voronoidis}.

\bibliography{defects_spacing_Bib}

\begin{thebibliography}{76}%
\makeatletter
\providecommand \@ifxundefined [1]{%
 \@ifx{#1\undefined}
}%
\providecommand \@ifnum [1]{%
 \ifnum #1\expandafter \@firstoftwo
 \else \expandafter \@secondoftwo
 \fi
}%
\providecommand \@ifx [1]{%
 \ifx #1\expandafter \@firstoftwo
 \else \expandafter \@secondoftwo
 \fi
}%
\providecommand \natexlab [1]{#1}%
\providecommand \enquote  [1]{``#1''}%
\providecommand \bibnamefont  [1]{#1}%
\providecommand \bibfnamefont [1]{#1}%
\providecommand \citenamefont [1]{#1}%
\providecommand \href@noop [0]{\@secondoftwo}%
\providecommand \href [0]{\begingroup \@sanitize@url \@href}%
\providecommand \@href[1]{\@@startlink{#1}\@@href}%
\providecommand \@@href[1]{\endgroup#1\@@endlink}%
\providecommand \@sanitize@url [0]{\catcode `\\12\catcode `\$12\catcode
  `\&12\catcode `\#12\catcode `\^12\catcode `\_12\catcode `\%12\relax}%
\providecommand \@@startlink[1]{}%
\providecommand \@@endlink[0]{}%
\providecommand \url  [0]{\begingroup\@sanitize@url \@url }%
\providecommand \@url [1]{\endgroup\@href {#1}{\urlprefix }}%
\providecommand \urlprefix  [0]{URL }%
\providecommand \Eprint [0]{\href }%
\providecommand \doibase [0]{https://doi.org/}%
\providecommand \selectlanguage [0]{\@gobble}%
\providecommand \bibinfo  [0]{\@secondoftwo}%
\providecommand \bibfield  [0]{\@secondoftwo}%
\providecommand \translation [1]{[#1]}%
\providecommand \BibitemOpen [0]{}%
\providecommand \bibitemStop [0]{}%
\providecommand \bibitemNoStop [0]{.\EOS\space}%
\providecommand \EOS [0]{\spacefactor3000\relax}%
\providecommand \BibitemShut  [1]{\csname bibitem#1\endcsname}%
\let\auto@bib@innerbib\@empty
\bibitem [{\citenamefont {Ueda}(2010)}]{Ueda10}%
  \BibitemOpen
  \bibfield  {author} {\bibinfo {author} {\bibfnamefont {M.}~\bibnamefont
  {Ueda}},\ }\href {https://doi.org/10.1142/7216} {\emph {\bibinfo {title}
  {Fundamentals and New Frontiers of Bose-Einstein Condensation}}}\ (\bibinfo
  {publisher} {World Scientific},\ \bibinfo {year} {2010})\BibitemShut
  {NoStop}%
\bibitem [{\citenamefont {Weiler}\ \emph {et~al.}(2008)\citenamefont {Weiler},
  \citenamefont {Neely}, \citenamefont {Scherer}, \citenamefont {Bradley},
  \citenamefont {Davis},\ and\ \citenamefont {Anderson}}]{Weiler08}%
  \BibitemOpen
  \bibfield  {author} {\bibinfo {author} {\bibfnamefont {C.~N.}\ \bibnamefont
  {Weiler}}, \bibinfo {author} {\bibfnamefont {T.~W.}\ \bibnamefont {Neely}},
  \bibinfo {author} {\bibfnamefont {D.~R.}\ \bibnamefont {Scherer}}, \bibinfo
  {author} {\bibfnamefont {A.~S.}\ \bibnamefont {Bradley}}, \bibinfo {author}
  {\bibfnamefont {M.~J.}\ \bibnamefont {Davis}},\ and\ \bibinfo {author}
  {\bibfnamefont {B.~P.}\ \bibnamefont {Anderson}},\ }\bibfield  {title}
  {\bibinfo {title} {Spontaneous vortices in the formation of {B}ose-{E}instein
  condensates},\ }\href {https://doi.org/10.1038/nature07334} {\bibfield
  {journal} {\bibinfo  {journal} {Nature}\ }\textbf {\bibinfo {volume} {455}},\
  \bibinfo {pages} {948} (\bibinfo {year} {2008})}\BibitemShut {NoStop}%
\bibitem [{\citenamefont {Abo-Shaeer}\ \emph {et~al.}(2001)\citenamefont
  {Abo-Shaeer}, \citenamefont {Raman}, \citenamefont {Vogels},\ and\
  \citenamefont {Ketterle}}]{Ketterle01}%
  \BibitemOpen
  \bibfield  {author} {\bibinfo {author} {\bibfnamefont {J.~R.}\ \bibnamefont
  {Abo-Shaeer}}, \bibinfo {author} {\bibfnamefont {C.}~\bibnamefont {Raman}},
  \bibinfo {author} {\bibfnamefont {J.~M.}\ \bibnamefont {Vogels}},\ and\
  \bibinfo {author} {\bibfnamefont {W.}~\bibnamefont {Ketterle}},\ }\bibfield
  {title} {\bibinfo {title} {Observation of vortex lattices in bose-einstein
  condensates},\ }\href {https://doi.org/10.1126/science.1060182} {\bibfield
  {journal} {\bibinfo  {journal} {Science}\ }\textbf {\bibinfo {volume}
  {292}},\ \bibinfo {pages} {476} (\bibinfo {year} {2001})}\BibitemShut
  {NoStop}%
\bibitem [{\citenamefont {Anderson}(2010)}]{Anderson10}%
  \BibitemOpen
  \bibfield  {author} {\bibinfo {author} {\bibfnamefont {B.~P.}\ \bibnamefont
  {Anderson}},\ }\bibfield  {title} {\bibinfo {title} {Resource article:
  Experiments with vortices in superfluid atomic gases},\ }\href
  {https://doi.org/10.1007/s10909-010-0224-1} {\bibfield  {journal} {\bibinfo
  {journal} {Journal of Low Temperature Physics}\ }\textbf {\bibinfo {volume}
  {161}},\ \bibinfo {pages} {574} (\bibinfo {year} {2010})}\BibitemShut
  {NoStop}%
\bibitem [{\citenamefont {Wilson}\ \emph {et~al.}(2015)\citenamefont {Wilson},
  \citenamefont {Newman}, \citenamefont {Lowney},\ and\ \citenamefont
  {Anderson}}]{Wilson15}%
  \BibitemOpen
  \bibfield  {author} {\bibinfo {author} {\bibfnamefont {K.~E.}\ \bibnamefont
  {Wilson}}, \bibinfo {author} {\bibfnamefont {Z.~L.}\ \bibnamefont {Newman}},
  \bibinfo {author} {\bibfnamefont {J.~D.}\ \bibnamefont {Lowney}},\ and\
  \bibinfo {author} {\bibfnamefont {B.~P.}\ \bibnamefont {Anderson}},\
  }\bibfield  {title} {\bibinfo {title} {In situ imaging of vortices in
  bose-einstein condensates},\ }\href
  {https://doi.org/10.1103/PhysRevA.91.023621} {\bibfield  {journal} {\bibinfo
  {journal} {Phys. Rev. A}\ }\textbf {\bibinfo {volume} {91}},\ \bibinfo
  {pages} {023621} (\bibinfo {year} {2015})}\BibitemShut {NoStop}%
\bibitem [{\citenamefont {Metz}\ \emph {et~al.}(2021)\citenamefont {Metz},
  \citenamefont {Polo}, \citenamefont {Weber},\ and\ \citenamefont
  {Busch}}]{Metz21}%
  \BibitemOpen
  \bibfield  {author} {\bibinfo {author} {\bibfnamefont {F.}~\bibnamefont
  {Metz}}, \bibinfo {author} {\bibfnamefont {J.}~\bibnamefont {Polo}}, \bibinfo
  {author} {\bibfnamefont {N.}~\bibnamefont {Weber}},\ and\ \bibinfo {author}
  {\bibfnamefont {T.}~\bibnamefont {Busch}},\ }\bibfield  {title} {\bibinfo
  {title} {Deep-learning-based quantum vortex detection in atomic
  bose–einstein condensates},\ }\href
  {https://doi.org/10.1088/2632-2153/abea6a} {\bibfield  {journal} {\bibinfo
  {journal} {Machine Learning: Science and Technology}\ }\textbf {\bibinfo
  {volume} {2}},\ \bibinfo {pages} {035019} (\bibinfo {year}
  {2021})}\BibitemShut {NoStop}%
\bibitem [{\citenamefont {Kim}\ \emph {et~al.}(2023)\citenamefont {Kim},
  \citenamefont {Kwon}, \citenamefont {Rabga},\ and\ \citenamefont
  {Shin}}]{Kim23}%
  \BibitemOpen
  \bibfield  {author} {\bibinfo {author} {\bibfnamefont {M.}~\bibnamefont
  {Kim}}, \bibinfo {author} {\bibfnamefont {J.}~\bibnamefont {Kwon}}, \bibinfo
  {author} {\bibfnamefont {T.}~\bibnamefont {Rabga}},\ and\ \bibinfo {author}
  {\bibfnamefont {Y.}~\bibnamefont {Shin}},\ }\bibfield  {title} {\bibinfo
  {title} {Vortex detection in atomic bose-einstein condensates using neural
  networks trained on synthetic images},\ }\href
  {https://doi.org/10.1088/2632-2153/ad03ad} {\bibfield  {journal} {\bibinfo
  {journal} {Machine Learning: Science and Technology}\ }\textbf {\bibinfo
  {volume} {4}},\ \bibinfo {pages} {045017} (\bibinfo {year}
  {2023})}\BibitemShut {NoStop}%
\bibitem [{\citenamefont {Kibble}(1976)}]{Kibble76a}%
  \BibitemOpen
  \bibfield  {author} {\bibinfo {author} {\bibfnamefont {T.~W.~B.}\
  \bibnamefont {Kibble}},\ }\bibfield  {title} {\bibinfo {title} {Topology of
  cosmic domains and strings},\ }\href
  {http://stacks.iop.org/0305-4470/9/i=8/a=029} {\bibfield  {journal} {\bibinfo
   {journal} {\href{http://stacks.iop.org/0305-4470/9/i=8/a=029}{J. of Phys. A:
  Math. Gen.}}\ }\textbf {\bibinfo {volume} {9}},\ \bibinfo {pages} {1387}
  (\bibinfo {year} {1976})}\BibitemShut {NoStop}%
\bibitem [{\citenamefont {Kibble}(1980)}]{Kibble76b}%
  \BibitemOpen
  \bibfield  {author} {\bibinfo {author} {\bibfnamefont {T.~W.~B.}\
  \bibnamefont {Kibble}},\ }\bibfield  {title} {\bibinfo {title} {Some
  implications of a cosmological phase transition},\ }\href
  {http://www.sciencedirect.com/science/article/pii/0370157380900915}
  {\bibfield  {journal} {\bibinfo  {journal}
  {\href{http://www.sciencedirect.com/science/article/pii/0370157380900915}{Phys.
  Reports}}\ }\textbf {\bibinfo {volume} {67}},\ \bibinfo {pages} {183}
  (\bibinfo {year} {1980})}\BibitemShut {NoStop}%
\bibitem [{\citenamefont {Zurek}(1985)}]{Zurek96a}%
  \BibitemOpen
  \bibfield  {author} {\bibinfo {author} {\bibfnamefont {W.~H.}\ \bibnamefont
  {Zurek}},\ }\bibfield  {title} {\bibinfo {title} {Cosmological experiments in
  superfluid helium?},\ }\href {http://dx.doi.org/10.1038/317505a0} {\bibfield
  {journal} {\bibinfo  {journal}
  {\href{http://dx.doi.org/10.1038/317505a0}{Nature}}\ }\textbf {\bibinfo
  {volume} {317}},\ \bibinfo {pages} {505} (\bibinfo {year}
  {1985})}\BibitemShut {NoStop}%
\bibitem [{\citenamefont {Zurek}(1993)}]{Zurek96b}%
  \BibitemOpen
  \bibfield  {author} {\bibinfo {author} {\bibfnamefont {W.~H.}\ \bibnamefont
  {Zurek}},\ }\bibfield  {title} {\bibinfo {title} {Cosmic strings in
  laboratory superfluids and the topological remnants of other phase
  transitions},\ }\href
  {http://www.actaphys.uj.edu.pl/fulltext?series=Reg&vol=24&page=1301}
  {\bibfield  {journal} {\bibinfo  {journal}
  {\href{http://www.actaphys.uj.edu.pl/fulltext?series=Reg&vol=24&page=1301}{Acta
  Phys. Pol. B}}\ }\textbf {\bibinfo {volume} {24}},\ \bibinfo {pages} {1301}
  (\bibinfo {year} {1993})}\BibitemShut {NoStop}%
\bibitem [{\citenamefont {del Campo}\ and\ \citenamefont {Zurek}(2014)}]{DZ14}%
  \BibitemOpen
  \bibfield  {author} {\bibinfo {author} {\bibfnamefont {A.}~\bibnamefont {del
  Campo}}\ and\ \bibinfo {author} {\bibfnamefont {W.~H.}\ \bibnamefont
  {Zurek}},\ }\bibfield  {title} {\bibinfo {title} {Universality of phase
  transition dynamics: Topological defects from symmetry breaking},\ }\href
  {https://doi.org/10.1142/S0217751X1430018X} {\bibfield  {journal} {\bibinfo
  {journal} {Int. J. Mod. Phys. A}\ }\textbf {\bibinfo {volume} {29}},\
  \bibinfo {pages} {1430018} (\bibinfo {year} {2014})}\BibitemShut {NoStop}%
\bibitem [{\citenamefont {Donner}\ \emph {et~al.}(2007)\citenamefont {Donner},
  \citenamefont {Ritter}, \citenamefont {Bourdel}, \citenamefont {{\"O}ttl},
  \citenamefont {K{\"o}hl},\ and\ \citenamefont {Esslinger}}]{Donner07}%
  \BibitemOpen
  \bibfield  {author} {\bibinfo {author} {\bibfnamefont {T.}~\bibnamefont
  {Donner}}, \bibinfo {author} {\bibfnamefont {S.}~\bibnamefont {Ritter}},
  \bibinfo {author} {\bibfnamefont {T.}~\bibnamefont {Bourdel}}, \bibinfo
  {author} {\bibfnamefont {A.}~\bibnamefont {{\"O}ttl}}, \bibinfo {author}
  {\bibfnamefont {M.}~\bibnamefont {K{\"o}hl}},\ and\ \bibinfo {author}
  {\bibfnamefont {T.}~\bibnamefont {Esslinger}},\ }\bibfield  {title} {\bibinfo
  {title} {Critical behavior of a trapped interacting bose gas},\ }\href
  {https://doi.org/10.1126/science.1138807} {\bibfield  {journal} {\bibinfo
  {journal} {Science}\ }\textbf {\bibinfo {volume} {315}},\ \bibinfo {pages}
  {1556} (\bibinfo {year} {2007})}\BibitemShut {NoStop}%
\bibitem [{\citenamefont {Campostrini}\ \emph {et~al.}(2001)\citenamefont
  {Campostrini}, \citenamefont {Hasenbusch}, \citenamefont {Pelissetto},
  \citenamefont {Rossi},\ and\ \citenamefont {Vicari}}]{Campostrini01}%
  \BibitemOpen
  \bibfield  {author} {\bibinfo {author} {\bibfnamefont {M.}~\bibnamefont
  {Campostrini}}, \bibinfo {author} {\bibfnamefont {M.}~\bibnamefont
  {Hasenbusch}}, \bibinfo {author} {\bibfnamefont {A.}~\bibnamefont
  {Pelissetto}}, \bibinfo {author} {\bibfnamefont {P.}~\bibnamefont {Rossi}},\
  and\ \bibinfo {author} {\bibfnamefont {E.}~\bibnamefont {Vicari}},\
  }\bibfield  {title} {\bibinfo {title} {Critical behavior of the
  three-dimensional $\mathrm{XY}$ universality class},\ }\href
  {https://doi.org/10.1103/PhysRevB.63.214503} {\bibfield  {journal} {\bibinfo
  {journal} {Phys. Rev. B}\ }\textbf {\bibinfo {volume} {63}},\ \bibinfo
  {pages} {214503} (\bibinfo {year} {2001})}\BibitemShut {NoStop}%
\bibitem [{\citenamefont {Navon}\ \emph {et~al.}(2015)\citenamefont {Navon},
  \citenamefont {Gaunt}, \citenamefont {Smith},\ and\ \citenamefont
  {Hadzibabic}}]{Navon15}%
  \BibitemOpen
  \bibfield  {author} {\bibinfo {author} {\bibfnamefont {N.}~\bibnamefont
  {Navon}}, \bibinfo {author} {\bibfnamefont {A.~L.}\ \bibnamefont {Gaunt}},
  \bibinfo {author} {\bibfnamefont {R.~P.}\ \bibnamefont {Smith}},\ and\
  \bibinfo {author} {\bibfnamefont {Z.}~\bibnamefont {Hadzibabic}},\ }\bibfield
   {title} {\bibinfo {title} {Critical dynamics of spontaneous symmetry
  breaking in a homogeneous {B}ose gas},\ }\href
  {https://doi.org/10.1126/science.1258676} {\bibfield  {journal} {\bibinfo
  {journal} {Science}\ }\textbf {\bibinfo {volume} {347}},\ \bibinfo {pages}
  {167} (\bibinfo {year} {2015})}\BibitemShut {NoStop}%
\bibitem [{\citenamefont {Hohenberg}\ and\ \citenamefont
  {Halperin}(1977)}]{Hohenberg_1977}%
  \BibitemOpen
  \bibfield  {author} {\bibinfo {author} {\bibfnamefont {P.~C.}\ \bibnamefont
  {Hohenberg}}\ and\ \bibinfo {author} {\bibfnamefont {B.~I.}\ \bibnamefont
  {Halperin}},\ }\bibfield  {title} {\bibinfo {title} {Theory of dynamic
  critical phenomena},\ }\href {https://doi.org/10.1103/RevModPhys.49.435}
  {\bibfield  {journal} {\bibinfo  {journal} {Rev. Mod. Phys.}\ }\textbf
  {\bibinfo {volume} {49}},\ \bibinfo {pages} {435} (\bibinfo {year}
  {1977})}\BibitemShut {NoStop}%
\bibitem [{\citenamefont {Chomaz}\ \emph {et~al.}(2015)\citenamefont {Chomaz},
  \citenamefont {Corman}, \citenamefont {Bienaim{\'e}}, \citenamefont
  {Desbuquois}, \citenamefont {Weitenberg}, \citenamefont {Nascimb{\`e}ne},
  \citenamefont {Beugnon},\ and\ \citenamefont {Dalibard}}]{Chomaz15}%
  \BibitemOpen
  \bibfield  {author} {\bibinfo {author} {\bibfnamefont {L.}~\bibnamefont
  {Chomaz}}, \bibinfo {author} {\bibfnamefont {L.}~\bibnamefont {Corman}},
  \bibinfo {author} {\bibfnamefont {T.}~\bibnamefont {Bienaim{\'e}}}, \bibinfo
  {author} {\bibfnamefont {R.}~\bibnamefont {Desbuquois}}, \bibinfo {author}
  {\bibfnamefont {C.}~\bibnamefont {Weitenberg}}, \bibinfo {author}
  {\bibfnamefont {S.}~\bibnamefont {Nascimb{\`e}ne}}, \bibinfo {author}
  {\bibfnamefont {J.}~\bibnamefont {Beugnon}},\ and\ \bibinfo {author}
  {\bibfnamefont {J.}~\bibnamefont {Dalibard}},\ }\bibfield  {title} {\bibinfo
  {title} {Emergence of coherence via transverse condensation in a uniform
  quasi-two-dimensional {B}ose gas},\ }\href
  {https://doi.org/10.1038/ncomms7162} {\bibfield  {journal} {\bibinfo
  {journal} {Nat. Comm.}\ }\textbf {\bibinfo {volume} {6}},\ \bibinfo {pages}
  {6162} (\bibinfo {year} {2015})}\BibitemShut {NoStop}%
\bibitem [{\citenamefont {Donadello}\ \emph {et~al.}(2016)\citenamefont
  {Donadello}, \citenamefont {Serafini}, \citenamefont {Bienaim\'e},
  \citenamefont {Dalfovo}, \citenamefont {Lamporesi},\ and\ \citenamefont
  {Ferrari}}]{Donadello16}%
  \BibitemOpen
  \bibfield  {author} {\bibinfo {author} {\bibfnamefont {S.}~\bibnamefont
  {Donadello}}, \bibinfo {author} {\bibfnamefont {S.}~\bibnamefont {Serafini}},
  \bibinfo {author} {\bibfnamefont {T.}~\bibnamefont {Bienaim\'e}}, \bibinfo
  {author} {\bibfnamefont {F.}~\bibnamefont {Dalfovo}}, \bibinfo {author}
  {\bibfnamefont {G.}~\bibnamefont {Lamporesi}},\ and\ \bibinfo {author}
  {\bibfnamefont {G.}~\bibnamefont {Ferrari}},\ }\bibfield  {title} {\bibinfo
  {title} {Creation and counting of defects in a temperature-quenched
  bose-einstein condensate},\ }\href
  {https://doi.org/10.1103/PhysRevA.94.023628} {\bibfield  {journal} {\bibinfo
  {journal} {Phys. Rev. A}\ }\textbf {\bibinfo {volume} {94}},\ \bibinfo
  {pages} {023628} (\bibinfo {year} {2016})}\BibitemShut {NoStop}%
\bibitem [{\citenamefont {Goo}\ \emph {et~al.}(2021)\citenamefont {Goo},
  \citenamefont {Lim},\ and\ \citenamefont {Shin}}]{Goo21}%
  \BibitemOpen
  \bibfield  {author} {\bibinfo {author} {\bibfnamefont {J.}~\bibnamefont
  {Goo}}, \bibinfo {author} {\bibfnamefont {Y.}~\bibnamefont {Lim}},\ and\
  \bibinfo {author} {\bibfnamefont {Y.}~\bibnamefont {Shin}},\ }\bibfield
  {title} {\bibinfo {title} {Defect saturation in a rapidly quenched {B}ose
  gas},\ }\href {https://doi.org/10.1103/PhysRevLett.127.115701} {\bibfield
  {journal} {\bibinfo  {journal} {Phys. Rev. Lett.}\ }\textbf {\bibinfo
  {volume} {127}},\ \bibinfo {pages} {115701} (\bibinfo {year}
  {2021})}\BibitemShut {NoStop}%
\bibitem [{\citenamefont {Rabga}\ \emph {et~al.}(2023)\citenamefont {Rabga},
  \citenamefont {Lee}, \citenamefont {Bae}, \citenamefont {Kim},\ and\
  \citenamefont {Shin}}]{Rabga23}%
  \BibitemOpen
  \bibfield  {author} {\bibinfo {author} {\bibfnamefont {T.}~\bibnamefont
  {Rabga}}, \bibinfo {author} {\bibfnamefont {Y.}~\bibnamefont {Lee}}, \bibinfo
  {author} {\bibfnamefont {D.}~\bibnamefont {Bae}}, \bibinfo {author}
  {\bibfnamefont {M.}~\bibnamefont {Kim}},\ and\ \bibinfo {author}
  {\bibfnamefont {Y.}~\bibnamefont {Shin}},\ }\bibfield  {title} {\bibinfo
  {title} {Variations of the kibble-zurek scaling exponents of trapped bose
  gases},\ }\href {https://doi.org/10.1103/PhysRevA.108.023315} {\bibfield
  {journal} {\bibinfo  {journal} {Phys. Rev. A}\ }\textbf {\bibinfo {volume}
  {108}},\ \bibinfo {pages} {023315} (\bibinfo {year} {2023})}\BibitemShut
  {NoStop}%
\bibitem [{\citenamefont {Ko}\ \emph {et~al.}(2019)\citenamefont {Ko},
  \citenamefont {Park},\ and\ \citenamefont {Shin}}]{Shin19}%
  \BibitemOpen
  \bibfield  {author} {\bibinfo {author} {\bibfnamefont {B.}~\bibnamefont
  {Ko}}, \bibinfo {author} {\bibfnamefont {J.~W.}\ \bibnamefont {Park}},\ and\
  \bibinfo {author} {\bibfnamefont {Y.}~\bibnamefont {Shin}},\ }\bibfield
  {title} {\bibinfo {title} {Kibble--zurek universality in a strongly
  interacting fermi superfluid},\ }\href
  {https://doi.org/10.1038/s41567-019-0650-1} {\bibfield  {journal} {\bibinfo
  {journal} {Nature Physics}\ }\textbf {\bibinfo {volume} {15}},\ \bibinfo
  {pages} {1227} (\bibinfo {year} {2019})}\BibitemShut {NoStop}%
\bibitem [{\citenamefont {Lee}\ \emph {et~al.}(2023)\citenamefont {Lee},
  \citenamefont {Kim}, \citenamefont {Kim},\ and\ \citenamefont
  {il~Shin}}]{lee2023observation}%
  \BibitemOpen
  \bibfield  {author} {\bibinfo {author} {\bibfnamefont {K.}~\bibnamefont
  {Lee}}, \bibinfo {author} {\bibfnamefont {S.}~\bibnamefont {Kim}}, \bibinfo
  {author} {\bibfnamefont {T.}~\bibnamefont {Kim}},\ and\ \bibinfo {author}
  {\bibfnamefont {Y.}~\bibnamefont {il~Shin}},\ }\href@noop {} {\bibinfo
  {title} {Observation of universal kibble-zurek scaling in an atomic fermi
  superfluid}} (\bibinfo {year} {2023}),\ \Eprint
  {https://arxiv.org/abs/2310.05437} {arXiv:2310.05437 [cond-mat.quant-gas]}
  \BibitemShut {NoStop}%
\bibitem [{\citenamefont {Kibble}\ and\ \citenamefont {Volovik}(1997)}]{KV97}%
  \BibitemOpen
  \bibfield  {author} {\bibinfo {author} {\bibfnamefont {T.~W.~B.}\
  \bibnamefont {Kibble}}\ and\ \bibinfo {author} {\bibfnamefont {G.~E.}\
  \bibnamefont {Volovik}},\ }\bibfield  {title} {\bibinfo {title} {On phase
  ordering behind the propagating front of a second-order transition},\ }\href
  {https://doi.org/10.1134/1.567332} {\bibfield  {journal} {\bibinfo  {journal}
  {\href{https://doi.org/10.1134/1.567332}{J. E. Theo. Phys. Letters}}\
  }\textbf {\bibinfo {volume} {65}},\ \bibinfo {pages} {102} (\bibinfo {year}
  {1997})}\BibitemShut {NoStop}%
\bibitem [{\citenamefont {Zurek}(2009)}]{Zurek09}%
  \BibitemOpen
  \bibfield  {author} {\bibinfo {author} {\bibfnamefont {W.~H.}\ \bibnamefont
  {Zurek}},\ }\bibfield  {title} {\bibinfo {title} {Causality in condensates:
  Gray solitons as relics of {BEC} formation},\ }\href
  {https://doi.org/10.1103/PhysRevLett.102.105702} {\bibfield  {journal}
  {\bibinfo  {journal}
  {\href{https://link.aps.org/doi/10.1103/PhysRevLett.102.105702}{Phys. Rev.
  Lett.}}\ }\textbf {\bibinfo {volume} {102}},\ \bibinfo {pages} {105702}
  (\bibinfo {year} {2009})}\BibitemShut {NoStop}%
\bibitem [{\citenamefont {del Campo}\ \emph {et~al.}(2011)\citenamefont {del
  Campo}, \citenamefont {Retzker},\ and\ \citenamefont {Plenio}}]{DRP11}%
  \BibitemOpen
  \bibfield  {author} {\bibinfo {author} {\bibfnamefont {A.}~\bibnamefont {del
  Campo}}, \bibinfo {author} {\bibfnamefont {A.}~\bibnamefont {Retzker}},\ and\
  \bibinfo {author} {\bibfnamefont {M.~B.}\ \bibnamefont {Plenio}},\ }\bibfield
   {title} {\bibinfo {title} {The inhomogeneous {K}ibble-{Z}urek mechanism:
  vortex nucleation during {B}ose-{E}instein condensation},\ }\href
  {http://stacks.iop.org/1367-2630/13/i=8/a=083022} {\bibfield  {journal}
  {\bibinfo  {journal}
  {\href{http://stacks.iop.org/1367-2630/13/i=8/a=083022}{New J. Phys.}}\
  }\textbf {\bibinfo {volume} {13}},\ \bibinfo {pages} {083022} (\bibinfo
  {year} {2011})}\BibitemShut {NoStop}%
\bibitem [{\citenamefont {Nikoghosyan}\ \emph {et~al.}(2016)\citenamefont
  {Nikoghosyan}, \citenamefont {Nigmatullin},\ and\ \citenamefont
  {Plenio}}]{Nikoghosyan16}%
  \BibitemOpen
  \bibfield  {author} {\bibinfo {author} {\bibfnamefont {G.}~\bibnamefont
  {Nikoghosyan}}, \bibinfo {author} {\bibfnamefont {R.}~\bibnamefont
  {Nigmatullin}},\ and\ \bibinfo {author} {\bibfnamefont {M.~B.}\ \bibnamefont
  {Plenio}},\ }\bibfield  {title} {\bibinfo {title} {Universality in the
  dynamics of second-order phase transitions},\ }\href
  {https://doi.org/10.1103/PhysRevLett.116.080601} {\bibfield  {journal}
  {\bibinfo  {journal} {Phys. Rev. Lett.}\ }\textbf {\bibinfo {volume} {116}},\
  \bibinfo {pages} {080601} (\bibinfo {year} {2016})}\BibitemShut {NoStop}%
\bibitem [{\citenamefont {del Campo}\ \emph {et~al.}(2013)\citenamefont {del
  Campo}, \citenamefont {Kibble},\ and\ \citenamefont {Zurek}}]{DKZ13}%
  \BibitemOpen
  \bibfield  {author} {\bibinfo {author} {\bibfnamefont {A.}~\bibnamefont {del
  Campo}}, \bibinfo {author} {\bibfnamefont {T.~W.~B.}\ \bibnamefont
  {Kibble}},\ and\ \bibinfo {author} {\bibfnamefont {W.~H.}\ \bibnamefont
  {Zurek}},\ }\bibfield  {title} {\bibinfo {title} {Causality and
  non-equilibrium second-order phase transitions in inhomogeneous systems},\
  }\href {https://doi.org/10.1088/0953-8984/25/40/404210} {\bibfield  {journal}
  {\bibinfo  {journal} {Journal of Physics: Condensed Matter}\ }\textbf
  {\bibinfo {volume} {25}},\ \bibinfo {pages} {404210} (\bibinfo {year}
  {2013})}\BibitemShut {NoStop}%
\bibitem [{\citenamefont {Ulm}\ \emph {et~al.}(2013)\citenamefont {Ulm},
  \citenamefont {Ro{\ss}nagel}, \citenamefont {Jacob}, \citenamefont
  {Deg{\"u}nther}, \citenamefont {Dawkins}, \citenamefont {Poschinger},
  \citenamefont {Nigmatullin}, \citenamefont {Retzker}, \citenamefont {Plenio},
  \citenamefont {Schmidt-Kaler},\ and\ \citenamefont {Singer}}]{Ulm13}%
  \BibitemOpen
  \bibfield  {author} {\bibinfo {author} {\bibfnamefont {S.}~\bibnamefont
  {Ulm}}, \bibinfo {author} {\bibfnamefont {J.}~\bibnamefont {Ro{\ss}nagel}},
  \bibinfo {author} {\bibfnamefont {G.}~\bibnamefont {Jacob}}, \bibinfo
  {author} {\bibfnamefont {C.}~\bibnamefont {Deg{\"u}nther}}, \bibinfo {author}
  {\bibfnamefont {S.~T.}\ \bibnamefont {Dawkins}}, \bibinfo {author}
  {\bibfnamefont {U.~G.}\ \bibnamefont {Poschinger}}, \bibinfo {author}
  {\bibfnamefont {R.}~\bibnamefont {Nigmatullin}}, \bibinfo {author}
  {\bibfnamefont {A.}~\bibnamefont {Retzker}}, \bibinfo {author} {\bibfnamefont
  {M.~B.}\ \bibnamefont {Plenio}}, \bibinfo {author} {\bibfnamefont
  {F.}~\bibnamefont {Schmidt-Kaler}},\ and\ \bibinfo {author} {\bibfnamefont
  {K.}~\bibnamefont {Singer}},\ }\bibfield  {title} {\bibinfo {title}
  {Observation of the {K}ibble-{Z}urek scaling law for defect formation in ion
  crystals},\ }\href {http://dx.doi.org/10.1038/ncomms3290} {\bibfield
  {journal} {\bibinfo  {journal} {Nat. Comm.}\ }\textbf {\bibinfo {volume}
  {4}},\ \bibinfo {pages} {2290} (\bibinfo {year} {2013})}\BibitemShut
  {NoStop}%
\bibitem [{\citenamefont {Pyka}\ \emph {et~al.}(2013)\citenamefont {Pyka},
  \citenamefont {Keller}, \citenamefont {Partner}, \citenamefont {Nigmatullin},
  \citenamefont {Burgermeister}, \citenamefont {Meier}, \citenamefont
  {Kuhlmann}, \citenamefont {Retzker}, \citenamefont {Plenio}, \citenamefont
  {Zurek}, \citenamefont {del Campo},\ and\ \citenamefont
  {Mehlst\"aubler}}]{Pyka13}%
  \BibitemOpen
  \bibfield  {author} {\bibinfo {author} {\bibfnamefont {K.}~\bibnamefont
  {Pyka}}, \bibinfo {author} {\bibfnamefont {J.}~\bibnamefont {Keller}},
  \bibinfo {author} {\bibfnamefont {H.~L.}\ \bibnamefont {Partner}}, \bibinfo
  {author} {\bibfnamefont {R.}~\bibnamefont {Nigmatullin}}, \bibinfo {author}
  {\bibfnamefont {T.}~\bibnamefont {Burgermeister}}, \bibinfo {author}
  {\bibfnamefont {D.~M.}\ \bibnamefont {Meier}}, \bibinfo {author}
  {\bibfnamefont {K.}~\bibnamefont {Kuhlmann}}, \bibinfo {author}
  {\bibfnamefont {A.}~\bibnamefont {Retzker}}, \bibinfo {author} {\bibfnamefont
  {M.~B.}\ \bibnamefont {Plenio}}, \bibinfo {author} {\bibfnamefont {W.~H.}\
  \bibnamefont {Zurek}}, \bibinfo {author} {\bibfnamefont {A.}~\bibnamefont
  {del Campo}},\ and\ \bibinfo {author} {\bibfnamefont {T.~E.}\ \bibnamefont
  {Mehlst\"aubler}},\ }\bibfield  {title} {\bibinfo {title} {Topological defect
  formation and spontaneous symmetry breaking in ion {C}oulomb crystals},\
  }\href {http://dx.doi.org/10.1038/ncomms3291} {\bibfield  {journal} {\bibinfo
   {journal} {\href{http://dx.doi.org/10.1038/ncomms3291}{Nat. Comm.}}\
  }\textbf {\bibinfo {volume} {4}},\ \bibinfo {pages} {2291} (\bibinfo {year}
  {2013})}\BibitemShut {NoStop}%
\bibitem [{\citenamefont {Kim}\ \emph {et~al.}(2022)\citenamefont {Kim},
  \citenamefont {Rabga}, \citenamefont {Lee}, \citenamefont {Goo},
  \citenamefont {Bae},\ and\ \citenamefont {Shin}}]{KimShin22}%
  \BibitemOpen
  \bibfield  {author} {\bibinfo {author} {\bibfnamefont {M.}~\bibnamefont
  {Kim}}, \bibinfo {author} {\bibfnamefont {T.}~\bibnamefont {Rabga}}, \bibinfo
  {author} {\bibfnamefont {Y.}~\bibnamefont {Lee}}, \bibinfo {author}
  {\bibfnamefont {J.}~\bibnamefont {Goo}}, \bibinfo {author} {\bibfnamefont
  {D.}~\bibnamefont {Bae}},\ and\ \bibinfo {author} {\bibfnamefont
  {Y.}~\bibnamefont {Shin}},\ }\bibfield  {title} {\bibinfo {title}
  {Suppression of spontaneous defect formation in inhomogeneous bose gases},\
  }\href {https://doi.org/10.1103/PhysRevA.106.L061301} {\bibfield  {journal}
  {\bibinfo  {journal} {Phys. Rev. A}\ }\textbf {\bibinfo {volume} {106}},\
  \bibinfo {pages} {L061301} (\bibinfo {year} {2022})}\BibitemShut {NoStop}%
\bibitem [{\citenamefont {del Campo}\ \emph {et~al.}(2010)\citenamefont {del
  Campo}, \citenamefont {De~Chiara}, \citenamefont {Morigi}, \citenamefont
  {Plenio},\ and\ \citenamefont {Retzker}}]{delcampo10}%
  \BibitemOpen
  \bibfield  {author} {\bibinfo {author} {\bibfnamefont {A.}~\bibnamefont {del
  Campo}}, \bibinfo {author} {\bibfnamefont {G.}~\bibnamefont {De~Chiara}},
  \bibinfo {author} {\bibfnamefont {G.}~\bibnamefont {Morigi}}, \bibinfo
  {author} {\bibfnamefont {M.~B.}\ \bibnamefont {Plenio}},\ and\ \bibinfo
  {author} {\bibfnamefont {A.}~\bibnamefont {Retzker}},\ }\bibfield  {title}
  {\bibinfo {title} {Structural defects in ion chains by quenching the external
  potential: The inhomogeneous {K}ibble-{Z}urek mechanism},\ }\href
  {https://doi.org/10.1103/PhysRevLett.105.075701} {\bibfield  {journal}
  {\bibinfo  {journal} {Phys. Rev. Lett.}\ }\textbf {\bibinfo {volume} {105}},\
  \bibinfo {pages} {075701} (\bibinfo {year} {2010})}\BibitemShut {NoStop}%
\bibitem [{\citenamefont {Chesler}\ \emph {et~al.}(2015)\citenamefont
  {Chesler}, \citenamefont {Garcia-Garcia},\ and\ \citenamefont
  {Liu}}]{Chesler:2014gya}%
  \BibitemOpen
  \bibfield  {author} {\bibinfo {author} {\bibfnamefont {P.~M.}\ \bibnamefont
  {Chesler}}, \bibinfo {author} {\bibfnamefont {A.~M.}\ \bibnamefont
  {Garcia-Garcia}},\ and\ \bibinfo {author} {\bibfnamefont {H.}~\bibnamefont
  {Liu}},\ }\bibfield  {title} {\bibinfo {title} {{Defect Formation beyond
  {K}ibble-{Z}urek Mechanism and Holography}},\ }\href
  {https://doi.org/10.1103/PhysRevX.5.021015} {\bibfield  {journal} {\bibinfo
  {journal} {Phys. Rev. X}\ }\textbf {\bibinfo {volume} {5}},\ \bibinfo {pages}
  {021015} (\bibinfo {year} {2015})}\BibitemShut {NoStop}%
\bibitem [{\citenamefont {Zeng}\ \emph {et~al.}(2023)\citenamefont {Zeng},
  \citenamefont {Xia},\ and\ \citenamefont {del Campo}}]{Zeng23}%
  \BibitemOpen
  \bibfield  {author} {\bibinfo {author} {\bibfnamefont {H.-B.}\ \bibnamefont
  {Zeng}}, \bibinfo {author} {\bibfnamefont {C.-Y.}\ \bibnamefont {Xia}},\ and\
  \bibinfo {author} {\bibfnamefont {A.}~\bibnamefont {del Campo}},\ }\bibfield
  {title} {\bibinfo {title} {Universal breakdown of kibble-zurek scaling in
  fast quenches across a phase transition},\ }\href
  {https://doi.org/10.1103/PhysRevLett.130.060402} {\bibfield  {journal}
  {\bibinfo  {journal} {Phys. Rev. Lett.}\ }\textbf {\bibinfo {volume} {130}},\
  \bibinfo {pages} {060402} (\bibinfo {year} {2023})}\BibitemShut {NoStop}%
\bibitem [{\citenamefont {Gupta}\ \emph {et~al.}(2005)\citenamefont {Gupta},
  \citenamefont {Murch}, \citenamefont {Moore}, \citenamefont {Purdy},\ and\
  \citenamefont {Stamper-Kurn}}]{Gupta05}%
  \BibitemOpen
  \bibfield  {author} {\bibinfo {author} {\bibfnamefont {S.}~\bibnamefont
  {Gupta}}, \bibinfo {author} {\bibfnamefont {K.~W.}\ \bibnamefont {Murch}},
  \bibinfo {author} {\bibfnamefont {K.~L.}\ \bibnamefont {Moore}}, \bibinfo
  {author} {\bibfnamefont {T.~P.}\ \bibnamefont {Purdy}},\ and\ \bibinfo
  {author} {\bibfnamefont {D.~M.}\ \bibnamefont {Stamper-Kurn}},\ }\bibfield
  {title} {\bibinfo {title} {Bose-einstein condensation in a circular
  waveguide},\ }\href {https://doi.org/10.1103/PhysRevLett.95.143201}
  {\bibfield  {journal} {\bibinfo  {journal} {Phys. Rev. Lett.}\ }\textbf
  {\bibinfo {volume} {95}},\ \bibinfo {pages} {143201} (\bibinfo {year}
  {2005})}\BibitemShut {NoStop}%
\bibitem [{\citenamefont {Henderson}\ \emph {et~al.}(2009)\citenamefont
  {Henderson}, \citenamefont {Ryu}, \citenamefont {MacCormick},\ and\
  \citenamefont {Boshier}}]{Henderson09}%
  \BibitemOpen
  \bibfield  {author} {\bibinfo {author} {\bibfnamefont {K.}~\bibnamefont
  {Henderson}}, \bibinfo {author} {\bibfnamefont {C.}~\bibnamefont {Ryu}},
  \bibinfo {author} {\bibfnamefont {C.}~\bibnamefont {MacCormick}},\ and\
  \bibinfo {author} {\bibfnamefont {M.~G.}\ \bibnamefont {Boshier}},\
  }\bibfield  {title} {\bibinfo {title} {Experimental demonstration of painting
  arbitrary and dynamic potentials for bose–einstein condensates},\ }\href
  {https://doi.org/10.1088/1367-2630/11/4/043030} {\bibfield  {journal}
  {\bibinfo  {journal} {New Journal of Physics}\ }\textbf {\bibinfo {volume}
  {11}},\ \bibinfo {pages} {043030} (\bibinfo {year} {2009})}\BibitemShut
  {NoStop}%
\bibitem [{\citenamefont {Gaunt}\ \emph {et~al.}(2013)\citenamefont {Gaunt},
  \citenamefont {Schmidutz}, \citenamefont {Gotlibovych}, \citenamefont
  {Smith},\ and\ \citenamefont {Hadzibabic}}]{Gaunt13}%
  \BibitemOpen
  \bibfield  {author} {\bibinfo {author} {\bibfnamefont {A.~L.}\ \bibnamefont
  {Gaunt}}, \bibinfo {author} {\bibfnamefont {T.~F.}\ \bibnamefont
  {Schmidutz}}, \bibinfo {author} {\bibfnamefont {I.}~\bibnamefont
  {Gotlibovych}}, \bibinfo {author} {\bibfnamefont {R.~P.}\ \bibnamefont
  {Smith}},\ and\ \bibinfo {author} {\bibfnamefont {Z.}~\bibnamefont
  {Hadzibabic}},\ }\bibfield  {title} {\bibinfo {title} {Bose-einstein
  condensation of atoms in a uniform potential},\ }\href
  {https://doi.org/10.1103/PhysRevLett.110.200406} {\bibfield  {journal}
  {\bibinfo  {journal} {Phys. Rev. Lett.}\ }\textbf {\bibinfo {volume} {110}},\
  \bibinfo {pages} {200406} (\bibinfo {year} {2013})}\BibitemShut {NoStop}%
\bibitem [{\citenamefont {Navon}\ \emph {et~al.}(2016)\citenamefont {Navon},
  \citenamefont {Gaunt}, \citenamefont {Smith},\ and\ \citenamefont
  {Hadzibabic}}]{Navon16}%
  \BibitemOpen
  \bibfield  {author} {\bibinfo {author} {\bibfnamefont {N.}~\bibnamefont
  {Navon}}, \bibinfo {author} {\bibfnamefont {A.~L.}\ \bibnamefont {Gaunt}},
  \bibinfo {author} {\bibfnamefont {R.~P.}\ \bibnamefont {Smith}},\ and\
  \bibinfo {author} {\bibfnamefont {Z.}~\bibnamefont {Hadzibabic}},\ }\bibfield
   {title} {\bibinfo {title} {Emergence of a turbulent cascade in a quantum
  gas},\ }\href {https://doi.org/10.1038/nature20114} {\bibfield  {journal}
  {\bibinfo  {journal} {Nature}\ }\textbf {\bibinfo {volume} {539}},\ \bibinfo
  {pages} {72} (\bibinfo {year} {2016})}\BibitemShut {NoStop}%
\bibitem [{\citenamefont {Mukherjee}\ \emph {et~al.}(2017)\citenamefont
  {Mukherjee}, \citenamefont {Yan}, \citenamefont {Patel}, \citenamefont
  {Hadzibabic}, \citenamefont {Yefsah}, \citenamefont {Struck},\ and\
  \citenamefont {Zwierlein}}]{Mukherjee17}%
  \BibitemOpen
  \bibfield  {author} {\bibinfo {author} {\bibfnamefont {B.}~\bibnamefont
  {Mukherjee}}, \bibinfo {author} {\bibfnamefont {Z.}~\bibnamefont {Yan}},
  \bibinfo {author} {\bibfnamefont {P.~B.}\ \bibnamefont {Patel}}, \bibinfo
  {author} {\bibfnamefont {Z.}~\bibnamefont {Hadzibabic}}, \bibinfo {author}
  {\bibfnamefont {T.}~\bibnamefont {Yefsah}}, \bibinfo {author} {\bibfnamefont
  {J.}~\bibnamefont {Struck}},\ and\ \bibinfo {author} {\bibfnamefont {M.~W.}\
  \bibnamefont {Zwierlein}},\ }\bibfield  {title} {\bibinfo {title}
  {Homogeneous atomic fermi gases},\ }\href
  {https://doi.org/10.1103/PhysRevLett.118.123401} {\bibfield  {journal}
  {\bibinfo  {journal} {Phys. Rev. Lett.}\ }\textbf {\bibinfo {volume} {118}},\
  \bibinfo {pages} {123401} (\bibinfo {year} {2017})}\BibitemShut {NoStop}%
\bibitem [{\citenamefont {Gauthier}\ \emph {et~al.}(2019)\citenamefont
  {Gauthier}, \citenamefont {Reeves}, \citenamefont {Yu}, \citenamefont
  {Bradley}, \citenamefont {Baker}, \citenamefont {Bell}, \citenamefont
  {Rubinsztein-Dunlop}, \citenamefont {Davis},\ and\ \citenamefont
  {Neely}}]{Guillaume2019giant}%
  \BibitemOpen
  \bibfield  {author} {\bibinfo {author} {\bibfnamefont {G.}~\bibnamefont
  {Gauthier}}, \bibinfo {author} {\bibfnamefont {M.~T.}\ \bibnamefont
  {Reeves}}, \bibinfo {author} {\bibfnamefont {X.}~\bibnamefont {Yu}}, \bibinfo
  {author} {\bibfnamefont {A.~S.}\ \bibnamefont {Bradley}}, \bibinfo {author}
  {\bibfnamefont {M.~A.}\ \bibnamefont {Baker}}, \bibinfo {author}
  {\bibfnamefont {T.~A.}\ \bibnamefont {Bell}}, \bibinfo {author}
  {\bibfnamefont {H.}~\bibnamefont {Rubinsztein-Dunlop}}, \bibinfo {author}
  {\bibfnamefont {M.~J.}\ \bibnamefont {Davis}},\ and\ \bibinfo {author}
  {\bibfnamefont {T.~W.}\ \bibnamefont {Neely}},\ }\bibfield  {title} {\bibinfo
  {title} {Giant vortex clusters in a two-dimensional quantum fluid},\ }\href
  {https://doi.org/10.1126/science.aat5718} {\bibfield  {journal} {\bibinfo
  {journal} {Science}\ }\textbf {\bibinfo {volume} {364}},\ \bibinfo {pages}
  {1264} (\bibinfo {year} {2019})}\BibitemShut {NoStop}%
\bibitem [{\citenamefont {G\'omez-Ruiz}\ \emph {et~al.}(2020)\citenamefont
  {G\'omez-Ruiz}, \citenamefont {Mayo},\ and\ \citenamefont {del
  Campo}}]{GomezRuiz20}%
  \BibitemOpen
  \bibfield  {author} {\bibinfo {author} {\bibfnamefont {F.~J.}\ \bibnamefont
  {G\'omez-Ruiz}}, \bibinfo {author} {\bibfnamefont {J.~J.}\ \bibnamefont
  {Mayo}},\ and\ \bibinfo {author} {\bibfnamefont {A.}~\bibnamefont {del
  Campo}},\ }\bibfield  {title} {\bibinfo {title} {Full counting statistics of
  topological defects after crossing a phase transition},\ }\href
  {https://doi.org/10.1103/PhysRevLett.124.240602} {\bibfield  {journal}
  {\bibinfo  {journal} {Phys. Rev. Lett.}\ }\textbf {\bibinfo {volume} {124}},\
  \bibinfo {pages} {240602} (\bibinfo {year} {2020})}\BibitemShut {NoStop}%
\bibitem [{\citenamefont {Mayo}\ \emph {et~al.}(2021)\citenamefont {Mayo},
  \citenamefont {Fan}, \citenamefont {Chern},\ and\ \citenamefont {del
  Campo}}]{Mayo21}%
  \BibitemOpen
  \bibfield  {author} {\bibinfo {author} {\bibfnamefont {J.~J.}\ \bibnamefont
  {Mayo}}, \bibinfo {author} {\bibfnamefont {Z.}~\bibnamefont {Fan}}, \bibinfo
  {author} {\bibfnamefont {G.-W.}\ \bibnamefont {Chern}},\ and\ \bibinfo
  {author} {\bibfnamefont {A.}~\bibnamefont {del Campo}},\ }\bibfield  {title}
  {\bibinfo {title} {Distribution of kinks in an {I}sing ferromagnet after
  annealing and the generalized {K}ibble-{Z}urek mechanism},\ }\href
  {https://doi.org/10.1103/PhysRevResearch.3.033150} {\bibfield  {journal}
  {\bibinfo  {journal} {Phys. Rev. Research}\ }\textbf {\bibinfo {volume}
  {3}},\ \bibinfo {pages} {033150} (\bibinfo {year} {2021})}\BibitemShut
  {NoStop}%
\bibitem [{\citenamefont {del Campo}\ \emph {et~al.}(2021)\citenamefont {del
  Campo}, \citenamefont {G{\'o}mez-Ruiz}, \citenamefont {Li}, \citenamefont
  {Xia}, \citenamefont {Zeng},\ and\ \citenamefont {Zhang}}]{delcampo21}%
  \BibitemOpen
  \bibfield  {author} {\bibinfo {author} {\bibfnamefont {A.}~\bibnamefont {del
  Campo}}, \bibinfo {author} {\bibfnamefont {F.~J.}\ \bibnamefont
  {G{\'o}mez-Ruiz}}, \bibinfo {author} {\bibfnamefont {Z.-H.}\ \bibnamefont
  {Li}}, \bibinfo {author} {\bibfnamefont {C.-Y.}\ \bibnamefont {Xia}},
  \bibinfo {author} {\bibfnamefont {H.-B.}\ \bibnamefont {Zeng}},\ and\
  \bibinfo {author} {\bibfnamefont {H.-Q.}\ \bibnamefont {Zhang}},\ }\bibfield
  {title} {\bibinfo {title} {Universal statistics of vortices in a newborn
  holographic superconductor: beyond the kibble-zurek mechanism},\ }\href
  {https://doi.org/10.1007/JHEP06(2021)061} {\bibfield  {journal} {\bibinfo
  {journal} {Journal of High Energy Physics}\ }\textbf {\bibinfo {volume}
  {2021}},\ \bibinfo {pages} {61} (\bibinfo {year} {2021})}\BibitemShut
  {NoStop}%
\bibitem [{\citenamefont {Subires}\ \emph {et~al.}(2022)\citenamefont
  {Subires}, \citenamefont {G\'omez-Ruiz}, \citenamefont {Ruiz-Garc\'{\i}a},
  \citenamefont {Alonso},\ and\ \citenamefont {del Campo}}]{Subires22}%
  \BibitemOpen
  \bibfield  {author} {\bibinfo {author} {\bibfnamefont {D.}~\bibnamefont
  {Subires}}, \bibinfo {author} {\bibfnamefont {F.~J.}\ \bibnamefont
  {G\'omez-Ruiz}}, \bibinfo {author} {\bibfnamefont {A.}~\bibnamefont
  {Ruiz-Garc\'{\i}a}}, \bibinfo {author} {\bibfnamefont {D.}~\bibnamefont
  {Alonso}},\ and\ \bibinfo {author} {\bibfnamefont {A.}~\bibnamefont {del
  Campo}},\ }\bibfield  {title} {\bibinfo {title} {Benchmarking quantum
  annealing dynamics: The spin-vector langevin model},\ }\href
  {https://doi.org/10.1103/PhysRevResearch.4.023104} {\bibfield  {journal}
  {\bibinfo  {journal} {Phys. Rev. Res.}\ }\textbf {\bibinfo {volume} {4}},\
  \bibinfo {pages} {023104} (\bibinfo {year} {2022})}\BibitemShut {NoStop}%
\bibitem [{\citenamefont {G\'omez-Ruiz}\ \emph {et~al.}(2022)\citenamefont
  {G\'omez-Ruiz}, \citenamefont {Subires},\ and\ \citenamefont {del
  Campo}}]{GomezRuiz22}%
  \BibitemOpen
  \bibfield  {author} {\bibinfo {author} {\bibfnamefont {F.~J.}\ \bibnamefont
  {G\'omez-Ruiz}}, \bibinfo {author} {\bibfnamefont {D.}~\bibnamefont
  {Subires}},\ and\ \bibinfo {author} {\bibfnamefont {A.}~\bibnamefont {del
  Campo}},\ }\bibfield  {title} {\bibinfo {title} {Role of boundary conditions
  in the full counting statistics of topological defects after crossing a
  continuous phase transition},\ }\href
  {https://doi.org/10.1103/PhysRevB.106.134302} {\bibfield  {journal} {\bibinfo
   {journal} {Phys. Rev. B}\ }\textbf {\bibinfo {volume} {106}},\ \bibinfo
  {pages} {134302} (\bibinfo {year} {2022})}\BibitemShut {NoStop}%
\bibitem [{\citenamefont {Chiu}\ \emph {et~al.}(2013)\citenamefont {Chiu},
  \citenamefont {Stoyan}, \citenamefont {Kendall},\ and\ \citenamefont
  {Mecke}}]{Chiu2013stochastic}%
  \BibitemOpen
  \bibfield  {author} {\bibinfo {author} {\bibfnamefont {S.}~\bibnamefont
  {Chiu}}, \bibinfo {author} {\bibfnamefont {D.}~\bibnamefont {Stoyan}},
  \bibinfo {author} {\bibfnamefont {W.}~\bibnamefont {Kendall}},\ and\ \bibinfo
  {author} {\bibfnamefont {J.}~\bibnamefont {Mecke}},\ }\href@noop {} {\emph
  {\bibinfo {title} {Stochastic Geometry and Its Applications}}},\ Wiley Series
  in Probability and Statistics\ (\bibinfo  {publisher} {Wiley},\ \bibinfo
  {year} {2013})\BibitemShut {NoStop}%
\bibitem [{\citenamefont {Damski}\ and\ \citenamefont
  {Zurek}(2010{\natexlab{a}})}]{Damski_soliton_2010}%
  \BibitemOpen
  \bibfield  {author} {\bibinfo {author} {\bibfnamefont {B.}~\bibnamefont
  {Damski}}\ and\ \bibinfo {author} {\bibfnamefont {W.~H.}\ \bibnamefont
  {Zurek}},\ }\bibfield  {title} {\bibinfo {title} {Soliton creation during a
  bose-einstein condensation},\ }\href
  {https://doi.org/10.1103/PhysRevLett.104.160404} {\bibfield  {journal}
  {\bibinfo  {journal} {Phys. Rev. Lett.}\ }\textbf {\bibinfo {volume} {104}},\
  \bibinfo {pages} {160404} (\bibinfo {year} {2010}{\natexlab{a}})}\BibitemShut
  {NoStop}%
\bibitem [{\citenamefont {Liu}\ \emph {et~al.}(2020)\citenamefont {Liu},
  \citenamefont {Dziarmaga}, \citenamefont {Gou}, \citenamefont {Dalfovo},\
  and\ \citenamefont {Proukakis}}]{Liu_kibble_2020}%
  \BibitemOpen
  \bibfield  {author} {\bibinfo {author} {\bibfnamefont {I.-K.}\ \bibnamefont
  {Liu}}, \bibinfo {author} {\bibfnamefont {J.}~\bibnamefont {Dziarmaga}},
  \bibinfo {author} {\bibfnamefont {S.-C.}\ \bibnamefont {Gou}}, \bibinfo
  {author} {\bibfnamefont {F.}~\bibnamefont {Dalfovo}},\ and\ \bibinfo {author}
  {\bibfnamefont {N.~P.}\ \bibnamefont {Proukakis}},\ }\bibfield  {title}
  {\bibinfo {title} {Kibble-zurek dynamics in a trapped ultracold bose gas},\
  }\href {https://doi.org/10.1103/PhysRevResearch.2.033183} {\bibfield
  {journal} {\bibinfo  {journal} {Phys. Rev. Research}\ }\textbf {\bibinfo
  {volume} {2}},\ \bibinfo {pages} {033183} (\bibinfo {year}
  {2020})}\BibitemShut {NoStop}%
\bibitem [{\citenamefont {Cockburn}\ and\ \citenamefont
  {Proukakis}(2009)}]{cockburn2009stochastic}%
  \BibitemOpen
  \bibfield  {author} {\bibinfo {author} {\bibfnamefont {S.~P.}\ \bibnamefont
  {Cockburn}}\ and\ \bibinfo {author} {\bibfnamefont {N.~P.}\ \bibnamefont
  {Proukakis}},\ }\bibfield  {title} {\bibinfo {title} {The stochastic
  gross-pitaevskii equation and some applications},\ }\href
  {https://doi.org/10.1134/S1054660X09040057} {\bibfield  {journal} {\bibinfo
  {journal} {Laser Physics}\ }\textbf {\bibinfo {volume} {19}},\ \bibinfo
  {pages} {558} (\bibinfo {year} {2009})}\BibitemShut {NoStop}%
\bibitem [{\citenamefont {Blakie}\ \emph {et~al.}(2008)\citenamefont {Blakie},
  \citenamefont {Bradley}, \citenamefont {Davis}, \citenamefont {Ballagh},\
  and\ \citenamefont {Gardiner}}]{blakie2008dynamics}%
  \BibitemOpen
  \bibfield  {author} {\bibinfo {author} {\bibfnamefont {P.~B.}\ \bibnamefont
  {Blakie}}, \bibinfo {author} {\bibfnamefont {A.~S.}\ \bibnamefont {Bradley}},
  \bibinfo {author} {\bibfnamefont {M.~J.}\ \bibnamefont {Davis}}, \bibinfo
  {author} {\bibfnamefont {R.~J.}\ \bibnamefont {Ballagh}},\ and\ \bibinfo
  {author} {\bibfnamefont {C.~W.}\ \bibnamefont {Gardiner}},\ }\bibfield
  {title} {\bibinfo {title} {Dynamics and statistical mechanics of ultra-cold
  bose gases using c-field techniques},\ }\href
  {https://doi.org/10.1080/00018730802564254} {\bibfield  {journal} {\bibinfo
  {journal} {Advances in Physics}\ }\textbf {\bibinfo {volume} {57}},\ \bibinfo
  {pages} {363} (\bibinfo {year} {2008})}\BibitemShut {NoStop}%
\bibitem [{\citenamefont {Stoof}(1999)}]{stoof1999coherent}%
  \BibitemOpen
  \bibfield  {author} {\bibinfo {author} {\bibfnamefont {H.~T.~C.}\
  \bibnamefont {Stoof}},\ }\bibfield  {title} {\bibinfo {title} {Coherent
  versus incoherent dynamics during bose-einstein condensation in atomic
  gases},\ }\href {https://doi.org/10.1023/A:1021897703053} {\bibfield
  {journal} {\bibinfo  {journal} {Journal of Low Temperature Physics}\ }\textbf
  {\bibinfo {volume} {114}},\ \bibinfo {pages} {11} (\bibinfo {year}
  {1999})}\BibitemShut {NoStop}%
\bibitem [{\citenamefont {Stoof}\ and\ \citenamefont
  {Bijlsma}(2001)}]{stoof2001dynamics}%
  \BibitemOpen
  \bibfield  {author} {\bibinfo {author} {\bibfnamefont {H.~T.~C.}\
  \bibnamefont {Stoof}}\ and\ \bibinfo {author} {\bibfnamefont {M.~J.}\
  \bibnamefont {Bijlsma}},\ }\bibfield  {title} {\bibinfo {title} {Dynamics of
  fluctuating bose-einstein condensates},\ }\href
  {https://doi.org/10.1023/A:1017519118408} {\bibfield  {journal} {\bibinfo
  {journal} {Journal of Low Temperature Physics}\ }\textbf {\bibinfo {volume}
  {124}},\ \bibinfo {pages} {431} (\bibinfo {year} {2001})}\BibitemShut
  {NoStop}%
\bibitem [{\citenamefont {Choi}\ \emph {et~al.}(1998)\citenamefont {Choi},
  \citenamefont {Morgan},\ and\ \citenamefont
  {Burnett}}]{Choi:1998phenomenological}%
  \BibitemOpen
  \bibfield  {author} {\bibinfo {author} {\bibfnamefont {S.}~\bibnamefont
  {Choi}}, \bibinfo {author} {\bibfnamefont {S.~A.}\ \bibnamefont {Morgan}},\
  and\ \bibinfo {author} {\bibfnamefont {K.}~\bibnamefont {Burnett}},\
  }\bibfield  {title} {\bibinfo {title} {Phenomenological damping in trapped
  atomic bose-einstein condensates},\ }\href
  {https://doi.org/10.1103/PhysRevA.57.4057} {\bibfield  {journal} {\bibinfo
  {journal} {Phys. Rev. A}\ }\textbf {\bibinfo {volume} {57}},\ \bibinfo
  {pages} {4057} (\bibinfo {year} {1998})}\BibitemShut {NoStop}%
\bibitem [{\citenamefont {Dennis}\ \emph {et~al.}(2013)\citenamefont {Dennis},
  \citenamefont {Hope},\ and\ \citenamefont {Johnsson}}]{dennis2013xmds2}%
  \BibitemOpen
  \bibfield  {author} {\bibinfo {author} {\bibfnamefont {G.~R.}\ \bibnamefont
  {Dennis}}, \bibinfo {author} {\bibfnamefont {J.~J.}\ \bibnamefont {Hope}},\
  and\ \bibinfo {author} {\bibfnamefont {M.~T.}\ \bibnamefont {Johnsson}},\
  }\bibfield  {title} {\bibinfo {title} {Xmds2: Fast, scalable simulation of
  coupled stochastic partial differential equations},\ }\href
  {https://doi.org/https://doi.org/10.1016/j.cpc.2012.08.016} {\bibfield
  {journal} {\bibinfo  {journal} {Computer Physics Communications}\ }\textbf
  {\bibinfo {volume} {184}},\ \bibinfo {pages} {201} (\bibinfo {year}
  {2013})}\BibitemShut {NoStop}%
\bibitem [{\citenamefont {Scherer}\ \emph {et~al.}(2007)\citenamefont
  {Scherer}, \citenamefont {Weiler}, \citenamefont {Neely},\ and\ \citenamefont
  {Anderson}}]{Scherer07}%
  \BibitemOpen
  \bibfield  {author} {\bibinfo {author} {\bibfnamefont {D.~R.}\ \bibnamefont
  {Scherer}}, \bibinfo {author} {\bibfnamefont {C.~N.}\ \bibnamefont {Weiler}},
  \bibinfo {author} {\bibfnamefont {T.~W.}\ \bibnamefont {Neely}},\ and\
  \bibinfo {author} {\bibfnamefont {B.~P.}\ \bibnamefont {Anderson}},\
  }\bibfield  {title} {\bibinfo {title} {Vortex formation by merging of
  multiple trapped {B}ose-{E}instein condensates},\ }\href
  {https://doi.org/10.1103/PhysRevLett.98.110402} {\bibfield  {journal}
  {\bibinfo  {journal} {Phys. Rev. Lett.}\ }\textbf {\bibinfo {volume} {98}},\
  \bibinfo {pages} {110402} (\bibinfo {year} {2007})}\BibitemShut {NoStop}%
\bibitem [{\citenamefont {Yates}\ and\ \citenamefont
  {Zurek}(1998)}]{YatesPRL98}%
  \BibitemOpen
  \bibfield  {author} {\bibinfo {author} {\bibfnamefont {A.}~\bibnamefont
  {Yates}}\ and\ \bibinfo {author} {\bibfnamefont {W.~H.}\ \bibnamefont
  {Zurek}},\ }\bibfield  {title} {\bibinfo {title} {Vortex formation in two
  dimensions: When symmetry breaks, how big are the pieces?},\ }\href
  {https://doi.org/10.1103/PhysRevLett.80.5477} {\bibfield  {journal} {\bibinfo
   {journal} {Phys. Rev. Lett.}\ }\textbf {\bibinfo {volume} {80}},\ \bibinfo
  {pages} {5477} (\bibinfo {year} {1998})}\BibitemShut {NoStop}%
\bibitem [{\citenamefont {Damski}\ and\ \citenamefont
  {Zurek}(2010{\natexlab{b}})}]{Damski10}%
  \BibitemOpen
  \bibfield  {author} {\bibinfo {author} {\bibfnamefont {B.}~\bibnamefont
  {Damski}}\ and\ \bibinfo {author} {\bibfnamefont {W.~H.}\ \bibnamefont
  {Zurek}},\ }\bibfield  {title} {\bibinfo {title} {Soliton creation during a
  bose-einstein condensation},\ }\href
  {https://doi.org/10.1103/PhysRevLett.104.160404} {\bibfield  {journal}
  {\bibinfo  {journal} {Phys. Rev. Lett.}\ }\textbf {\bibinfo {volume} {104}},\
  \bibinfo {pages} {160404} (\bibinfo {year} {2010}{\natexlab{b}})}\BibitemShut
  {NoStop}%
\bibitem [{\citenamefont {Sonner}\ \emph {et~al.}(2015)\citenamefont {Sonner},
  \citenamefont {del Campo},\ and\ \citenamefont {Zurek}}]{Sonner15}%
  \BibitemOpen
  \bibfield  {author} {\bibinfo {author} {\bibfnamefont {J.}~\bibnamefont
  {Sonner}}, \bibinfo {author} {\bibfnamefont {A.}~\bibnamefont {del Campo}},\
  and\ \bibinfo {author} {\bibfnamefont {W.~H.}\ \bibnamefont {Zurek}},\
  }\bibfield  {title} {\bibinfo {title} {Universal far-from-equilibrium
  dynamics of a holographic superconductor},\ }\href
  {https://doi.org/10.1038/ncomms8406} {\bibfield  {journal} {\bibinfo
  {journal} {Nat. Comm.}\ }\textbf {\bibinfo {volume} {6}},\ \bibinfo {pages}
  {7406} (\bibinfo {year} {2015})}\BibitemShut {NoStop}%
\bibitem [{\citenamefont {Reichhardt}\ \emph {et~al.}(2022)\citenamefont
  {Reichhardt}, \citenamefont {del Campo},\ and\ \citenamefont
  {Reichhardt}}]{Reichhardt22}%
  \BibitemOpen
  \bibfield  {author} {\bibinfo {author} {\bibfnamefont {C.~J.~O.}\
  \bibnamefont {Reichhardt}}, \bibinfo {author} {\bibfnamefont
  {A.}~\bibnamefont {del Campo}},\ and\ \bibinfo {author} {\bibfnamefont
  {C.}~\bibnamefont {Reichhardt}},\ }\bibfield  {title} {\bibinfo {title}
  {Kibble-zurek mechanism for nonequilibrium phase transitions in driven
  systems with quenched disorder},\ }\href
  {https://doi.org/10.1038/s42005-022-00952-w} {\bibfield  {journal} {\bibinfo
  {journal} {Communications Physics}\ }\textbf {\bibinfo {volume} {5}},\
  \bibinfo {pages} {173} (\bibinfo {year} {2022})}\BibitemShut {NoStop}%
\bibitem [{\citenamefont {Mathai}(1999)}]{mathai1999introduction}%
  \BibitemOpen
  \bibfield  {author} {\bibinfo {author} {\bibfnamefont {A.~M.}\ \bibnamefont
  {Mathai}},\ }\href@noop {} {\emph {\bibinfo {title} {An introduction to
  geometrical probability: distributional aspects with applications}}},\
  Vol.~\bibinfo {volume} {1}\ (\bibinfo  {publisher} {CRC Press},\ \bibinfo
  {year} {1999})\BibitemShut {NoStop}%
\bibitem [{\citenamefont {del Campo}\ \emph {et~al.}(2022)\citenamefont {del
  Campo}, \citenamefont {G\'omez-Ruiz},\ and\ \citenamefont
  {Zhang}}]{delcampo22}%
  \BibitemOpen
  \bibfield  {author} {\bibinfo {author} {\bibfnamefont {A.}~\bibnamefont {del
  Campo}}, \bibinfo {author} {\bibfnamefont {F.~J.}\ \bibnamefont
  {G\'omez-Ruiz}},\ and\ \bibinfo {author} {\bibfnamefont {H.-Q.}\ \bibnamefont
  {Zhang}},\ }\bibfield  {title} {\bibinfo {title} {Locality of spontaneous
  symmetry breaking and universal spacing distribution of topological defects
  formed across a phase transition},\ }\href
  {https://doi.org/10.1103/PhysRevB.106.L140101} {\bibfield  {journal}
  {\bibinfo  {journal} {Phys. Rev. B}\ }\textbf {\bibinfo {volume} {106}},\
  \bibinfo {pages} {L140101} (\bibinfo {year} {2022})}\BibitemShut {NoStop}%
\bibitem [{\citenamefont {Lellouche}\ and\ \citenamefont
  {Souris}(2020)}]{lellouche2019distribution}%
  \BibitemOpen
  \bibfield  {author} {\bibinfo {author} {\bibfnamefont {S.}~\bibnamefont
  {Lellouche}}\ and\ \bibinfo {author} {\bibfnamefont {M.}~\bibnamefont
  {Souris}},\ }\bibfield  {title} {\bibinfo {title} {Distribution of distances
  between elements in a compact set},\ }\href
  {https://doi.org/10.3390/stats3010001} {\bibfield  {journal} {\bibinfo
  {journal} {Stats}\ }\textbf {\bibinfo {volume} {3}},\ \bibinfo {pages} {1}
  (\bibinfo {year} {2020})}\BibitemShut {NoStop}%
\bibitem [{\citenamefont {Nelson}(2002)}]{Nelson02book}%
  \BibitemOpen
  \bibfield  {author} {\bibinfo {author} {\bibfnamefont {D.~R.}\ \bibnamefont
  {Nelson}},\ }\href@noop {} {\emph {\bibinfo {title} {Defects and Geometry in
  {C}ondensed {M}atter {P}hysics}}}\ (\bibinfo  {publisher} {Cambridge
  University Press},\ \bibinfo {year} {2002})\BibitemShut {NoStop}%
\bibitem [{\citenamefont {Reichhardt}\ and\ \citenamefont
  {Olson~Reichhardt}(2003)}]{ReichhardtReichhardt03}%
  \BibitemOpen
  \bibfield  {author} {\bibinfo {author} {\bibfnamefont {C.}~\bibnamefont
  {Reichhardt}}\ and\ \bibinfo {author} {\bibfnamefont {C.~J.}\ \bibnamefont
  {Olson~Reichhardt}},\ }\bibfield  {title} {\bibinfo {title} {Fluctuating
  topological defects in 2d liquids: Heterogeneous motion and noise},\ }\href
  {https://doi.org/10.1103/PhysRevLett.90.095504} {\bibfield  {journal}
  {\bibinfo  {journal} {Phys. Rev. Lett.}\ }\textbf {\bibinfo {volume} {90}},\
  \bibinfo {pages} {095504} (\bibinfo {year} {2003})}\BibitemShut {NoStop}%
\bibitem [{\citenamefont {Jim\'enez}(2021)}]{Jimenez21}%
  \BibitemOpen
  \bibfield  {author} {\bibinfo {author} {\bibfnamefont {J.}~\bibnamefont
  {Jim\'enez}},\ }\bibfield  {title} {\bibinfo {title} {Collective organization
  and screening in two-dimensional turbulence},\ }\href
  {https://doi.org/10.1103/PhysRevFluids.6.084601} {\bibfield  {journal}
  {\bibinfo  {journal} {Phys. Rev. Fluids}\ }\textbf {\bibinfo {volume} {6}},\
  \bibinfo {pages} {084601} (\bibinfo {year} {2021})}\BibitemShut {NoStop}%
\bibitem [{\citenamefont {Stoop}\ and\ \citenamefont {Dunkel}(2018)}]{Stoop18}%
  \BibitemOpen
  \bibfield  {author} {\bibinfo {author} {\bibfnamefont {N.}~\bibnamefont
  {Stoop}}\ and\ \bibinfo {author} {\bibfnamefont {J.}~\bibnamefont {Dunkel}},\
  }\bibfield  {title} {\bibinfo {title} {Defect formation dynamics in curved
  elastic surface crystals},\ }\href {https://doi.org/10.1039/C7SM02233F}
  {\bibfield  {journal} {\bibinfo  {journal} {Soft Matter}\ }\textbf {\bibinfo
  {volume} {14}},\ \bibinfo {pages} {2329} (\bibinfo {year}
  {2018})}\BibitemShut {NoStop}%
\bibitem [{\citenamefont {Reichhardt}\ and\ \citenamefont
  {Reichhardt}(2023)}]{Reichhardt23}%
  \BibitemOpen
  \bibfield  {author} {\bibinfo {author} {\bibfnamefont {C.}~\bibnamefont
  {Reichhardt}}\ and\ \bibinfo {author} {\bibfnamefont {C.~J.~O.}\ \bibnamefont
  {Reichhardt}},\ }\bibfield  {title} {\bibinfo {title} {Kibble-zurek scenario
  and coarsening across nonequilibrium phase transitions in driven vortices and
  skyrmions},\ }\href {https://doi.org/10.1103/PhysRevResearch.5.033221}
  {\bibfield  {journal} {\bibinfo  {journal} {Phys. Rev. Res.}\ }\textbf
  {\bibinfo {volume} {5}},\ \bibinfo {pages} {033221} (\bibinfo {year}
  {2023})}\BibitemShut {NoStop}%
\bibitem [{\citenamefont {Weaire}\ \emph {et~al.}(1986)\citenamefont {Weaire},
  \citenamefont {Kermode},\ and\ \citenamefont {Wejchert}}]{Weaire:1986Weaire}%
  \BibitemOpen
  \bibfield  {author} {\bibinfo {author} {\bibfnamefont {D.}~\bibnamefont
  {Weaire}}, \bibinfo {author} {\bibfnamefont {J.~P.}\ \bibnamefont
  {Kermode}},\ and\ \bibinfo {author} {\bibfnamefont {J.}~\bibnamefont
  {Wejchert}},\ }\bibfield  {title} {\bibinfo {title} {On the distribution of
  cell areas in a voronoi network},\ }\href
  {https://doi.org/10.1080/13642818608240647} {\bibfield  {journal} {\bibinfo
  {journal} {Philosophical Magazine B}\ }\textbf {\bibinfo {volume} {53}},\
  \bibinfo {pages} {L101} (\bibinfo {year} {1986})}\BibitemShut {NoStop}%
\bibitem [{\citenamefont {Ferenc}\ and\ \citenamefont
  {N{\'e}da}(2007)}]{FERENC2007518}%
  \BibitemOpen
  \bibfield  {author} {\bibinfo {author} {\bibfnamefont {J.-S.}\ \bibnamefont
  {Ferenc}}\ and\ \bibinfo {author} {\bibfnamefont {Z.}~\bibnamefont
  {N{\'e}da}},\ }\bibfield  {title} {\bibinfo {title} {On the size distribution
  of poisson voronoi cells},\ }\href
  {https://doi.org/https://doi.org/10.1016/j.physa.2007.07.063} {\bibfield
  {journal} {\bibinfo  {journal} {Physica A: Statistical Mechanics and its
  Applications}\ }\textbf {\bibinfo {volume} {385}},\ \bibinfo {pages} {518}
  (\bibinfo {year} {2007})}\BibitemShut {NoStop}%
\bibitem [{\citenamefont {del Campo}(2018)}]{delcampo18}%
  \BibitemOpen
  \bibfield  {author} {\bibinfo {author} {\bibfnamefont {A.}~\bibnamefont {del
  Campo}},\ }\bibfield  {title} {\bibinfo {title} {Universal statistics of
  topological defects formed in a quantum phase transition},\ }\href
  {https://doi.org/10.1103/PhysRevLett.121.200601} {\bibfield  {journal}
  {\bibinfo  {journal} {Phys. Rev. Lett.}\ }\textbf {\bibinfo {volume} {121}},\
  \bibinfo {pages} {200601} (\bibinfo {year} {2018})}\BibitemShut {NoStop}%
\bibitem [{\citenamefont {Cui}\ \emph {et~al.}(2020)\citenamefont {Cui},
  \citenamefont {G{\'o}mez-Ruiz}, \citenamefont {Huang}, \citenamefont {Li},
  \citenamefont {Guo},\ and\ \citenamefont {del Campo}}]{Cui20}%
  \BibitemOpen
  \bibfield  {author} {\bibinfo {author} {\bibfnamefont {J.-M.}\ \bibnamefont
  {Cui}}, \bibinfo {author} {\bibfnamefont {F.~J.}\ \bibnamefont
  {G{\'o}mez-Ruiz}}, \bibinfo {author} {\bibfnamefont {Y.-F.}\ \bibnamefont
  {Huang}}, \bibinfo {author} {\bibfnamefont {C.-F.}\ \bibnamefont {Li}},
  \bibinfo {author} {\bibfnamefont {G.-C.}\ \bibnamefont {Guo}},\ and\ \bibinfo
  {author} {\bibfnamefont {A.}~\bibnamefont {del Campo}},\ }\bibfield  {title}
  {\bibinfo {title} {Experimentally testing quantum critical dynamics beyond
  the kibble--zurek mechanism},\ }\href
  {https://doi.org/10.1038/s42005-020-0306-6} {\bibfield  {journal} {\bibinfo
  {journal} {Communications Physics}\ }\textbf {\bibinfo {volume} {3}},\
  \bibinfo {pages} {44} (\bibinfo {year} {2020})}\BibitemShut {NoStop}%
\bibitem [{\citenamefont {Bando}\ \emph {et~al.}(2020)\citenamefont {Bando},
  \citenamefont {Susa}, \citenamefont {Oshiyama}, \citenamefont {Shibata},
  \citenamefont {Ohzeki}, \citenamefont {G\'omez-Ruiz}, \citenamefont {Lidar},
  \citenamefont {Suzuki}, \citenamefont {del Campo},\ and\ \citenamefont
  {Nishimori}}]{Bando20}%
  \BibitemOpen
  \bibfield  {author} {\bibinfo {author} {\bibfnamefont {Y.}~\bibnamefont
  {Bando}}, \bibinfo {author} {\bibfnamefont {Y.}~\bibnamefont {Susa}},
  \bibinfo {author} {\bibfnamefont {H.}~\bibnamefont {Oshiyama}}, \bibinfo
  {author} {\bibfnamefont {N.}~\bibnamefont {Shibata}}, \bibinfo {author}
  {\bibfnamefont {M.}~\bibnamefont {Ohzeki}}, \bibinfo {author} {\bibfnamefont
  {F.~J.}\ \bibnamefont {G\'omez-Ruiz}}, \bibinfo {author} {\bibfnamefont
  {D.~A.}\ \bibnamefont {Lidar}}, \bibinfo {author} {\bibfnamefont
  {S.}~\bibnamefont {Suzuki}}, \bibinfo {author} {\bibfnamefont
  {A.}~\bibnamefont {del Campo}},\ and\ \bibinfo {author} {\bibfnamefont
  {H.}~\bibnamefont {Nishimori}},\ }\bibfield  {title} {\bibinfo {title}
  {Probing the universality of topological defect formation in a quantum
  annealer: {K}ibble-{Z}urek mechanism and beyond},\ }\href
  {https://doi.org/10.1103/PhysRevResearch.2.033369} {\bibfield  {journal}
  {\bibinfo  {journal} {Phys. Rev. Research}\ }\textbf {\bibinfo {volume}
  {2}},\ \bibinfo {pages} {033369} (\bibinfo {year} {2020})}\BibitemShut
  {NoStop}%
\bibitem [{\citenamefont {King}\ \emph {et~al.}(2022)\citenamefont {King},
  \citenamefont {Suzuki}, \citenamefont {Raymond}, \citenamefont {Zucca},
  \citenamefont {Lanting}, \citenamefont {Altomare}, \citenamefont {Berkley},
  \citenamefont {Ejtemaee}, \citenamefont {Hoskinson}, \citenamefont {Huang},
  \citenamefont {Ladizinsky}, \citenamefont {MacDonald}, \citenamefont
  {Marsden}, \citenamefont {Oh}, \citenamefont {Poulin-Lamarre}, \citenamefont
  {Reis}, \citenamefont {Rich}, \citenamefont {Sato}, \citenamefont
  {Whittaker}, \citenamefont {Yao}, \citenamefont {Harris}, \citenamefont
  {Lidar}, \citenamefont {Nishimori},\ and\ \citenamefont {Amin}}]{King22}%
  \BibitemOpen
  \bibfield  {author} {\bibinfo {author} {\bibfnamefont {A.~D.}\ \bibnamefont
  {King}}, \bibinfo {author} {\bibfnamefont {S.}~\bibnamefont {Suzuki}},
  \bibinfo {author} {\bibfnamefont {J.}~\bibnamefont {Raymond}}, \bibinfo
  {author} {\bibfnamefont {A.}~\bibnamefont {Zucca}}, \bibinfo {author}
  {\bibfnamefont {T.}~\bibnamefont {Lanting}}, \bibinfo {author} {\bibfnamefont
  {F.}~\bibnamefont {Altomare}}, \bibinfo {author} {\bibfnamefont {A.~J.}\
  \bibnamefont {Berkley}}, \bibinfo {author} {\bibfnamefont {S.}~\bibnamefont
  {Ejtemaee}}, \bibinfo {author} {\bibfnamefont {E.}~\bibnamefont {Hoskinson}},
  \bibinfo {author} {\bibfnamefont {S.}~\bibnamefont {Huang}}, \bibinfo
  {author} {\bibfnamefont {E.}~\bibnamefont {Ladizinsky}}, \bibinfo {author}
  {\bibfnamefont {A.~J.~R.}\ \bibnamefont {MacDonald}}, \bibinfo {author}
  {\bibfnamefont {G.}~\bibnamefont {Marsden}}, \bibinfo {author} {\bibfnamefont
  {T.}~\bibnamefont {Oh}}, \bibinfo {author} {\bibfnamefont {G.}~\bibnamefont
  {Poulin-Lamarre}}, \bibinfo {author} {\bibfnamefont {M.}~\bibnamefont
  {Reis}}, \bibinfo {author} {\bibfnamefont {C.}~\bibnamefont {Rich}}, \bibinfo
  {author} {\bibfnamefont {Y.}~\bibnamefont {Sato}}, \bibinfo {author}
  {\bibfnamefont {J.~D.}\ \bibnamefont {Whittaker}}, \bibinfo {author}
  {\bibfnamefont {J.}~\bibnamefont {Yao}}, \bibinfo {author} {\bibfnamefont
  {R.}~\bibnamefont {Harris}}, \bibinfo {author} {\bibfnamefont {D.~A.}\
  \bibnamefont {Lidar}}, \bibinfo {author} {\bibfnamefont {H.}~\bibnamefont
  {Nishimori}},\ and\ \bibinfo {author} {\bibfnamefont {M.~H.}\ \bibnamefont
  {Amin}},\ }\bibfield  {title} {\bibinfo {title} {Coherent quantum annealing
  in a programmable 2,000{\thinspace}qubit ising chain},\ }\href
  {https://doi.org/10.1038/s41567-022-01741-6} {\bibfield  {journal} {\bibinfo
  {journal} {Nature Physics}\ }\textbf {\bibinfo {volume} {18}},\ \bibinfo
  {pages} {1324} (\bibinfo {year} {2022})}\BibitemShut {NoStop}%
\bibitem [{\citenamefont {Gherardini}\ \emph {et~al.}(2023)\citenamefont
  {Gherardini}, \citenamefont {Buffoni},\ and\ \citenamefont
  {Defenu}}]{Gherardini23}%
  \BibitemOpen
  \bibfield  {author} {\bibinfo {author} {\bibfnamefont {S.}~\bibnamefont
  {Gherardini}}, \bibinfo {author} {\bibfnamefont {L.}~\bibnamefont
  {Buffoni}},\ and\ \bibinfo {author} {\bibfnamefont {N.}~\bibnamefont
  {Defenu}},\ }\href@noop {} {\bibinfo {title} {Universal defects statistics
  with strong long-range interactions}} (\bibinfo {year} {2023}),\ \Eprint
  {https://arxiv.org/abs/2305.11771} {arXiv:2305.11771 [quant-ph]} \BibitemShut
  {NoStop}%
\bibitem [{\citenamefont {Carusotto}\ and\ \citenamefont
  {Ciuti}(2013)}]{Carusotto13}%
  \BibitemOpen
  \bibfield  {author} {\bibinfo {author} {\bibfnamefont {I.}~\bibnamefont
  {Carusotto}}\ and\ \bibinfo {author} {\bibfnamefont {C.}~\bibnamefont
  {Ciuti}},\ }\bibfield  {title} {\bibinfo {title} {Quantum fluids of light},\
  }\href {https://doi.org/10.1103/RevModPhys.85.299} {\bibfield  {journal}
  {\bibinfo  {journal} {Rev. Mod. Phys.}\ }\textbf {\bibinfo {volume} {85}},\
  \bibinfo {pages} {299} (\bibinfo {year} {2013})}\BibitemShut {NoStop}%
\bibitem [{\citenamefont {Deutschl{\"a}nder}\ \emph {et~al.}(2015)\citenamefont
  {Deutschl{\"a}nder}, \citenamefont {Dillmann}, \citenamefont {Maret},\ and\
  \citenamefont {Keim}}]{Keim15}%
  \BibitemOpen
  \bibfield  {author} {\bibinfo {author} {\bibfnamefont {S.}~\bibnamefont
  {Deutschl{\"a}nder}}, \bibinfo {author} {\bibfnamefont {P.}~\bibnamefont
  {Dillmann}}, \bibinfo {author} {\bibfnamefont {G.}~\bibnamefont {Maret}},\
  and\ \bibinfo {author} {\bibfnamefont {P.}~\bibnamefont {Keim}},\ }\bibfield
  {title} {\bibinfo {title} {{K}ibble{\textendash}{Z}urek mechanism in
  colloidal monolayers},\ }\href {https://doi.org/10.1073/pnas.1500763112}
  {\bibfield  {journal} {\bibinfo  {journal} {Proc. Nat. Acad. Sciences}\
  }\textbf {\bibinfo {volume} {112}},\ \bibinfo {pages} {6925} (\bibinfo {year}
  {2015})}\BibitemShut {NoStop}%
\bibitem [{\citenamefont {Lin}\ \emph {et~al.}(2014)\citenamefont {Lin},
  \citenamefont {Wang}, \citenamefont {Kamiya}, \citenamefont {Chern},
  \citenamefont {Fan}, \citenamefont {Fan}, \citenamefont {Casas},
  \citenamefont {Liu}, \citenamefont {Kiryukhin}, \citenamefont {Zurek},
  \citenamefont {Batista},\ and\ \citenamefont {Cheong}}]{Lin14}%
  \BibitemOpen
  \bibfield  {author} {\bibinfo {author} {\bibfnamefont {S.-Z.}\ \bibnamefont
  {Lin}}, \bibinfo {author} {\bibfnamefont {X.}~\bibnamefont {Wang}}, \bibinfo
  {author} {\bibfnamefont {Y.}~\bibnamefont {Kamiya}}, \bibinfo {author}
  {\bibfnamefont {G.-W.}\ \bibnamefont {Chern}}, \bibinfo {author}
  {\bibfnamefont {F.}~\bibnamefont {Fan}}, \bibinfo {author} {\bibfnamefont
  {D.}~\bibnamefont {Fan}}, \bibinfo {author} {\bibfnamefont {B.}~\bibnamefont
  {Casas}}, \bibinfo {author} {\bibfnamefont {Y.}~\bibnamefont {Liu}}, \bibinfo
  {author} {\bibfnamefont {V.}~\bibnamefont {Kiryukhin}}, \bibinfo {author}
  {\bibfnamefont {W.~H.}\ \bibnamefont {Zurek}}, \bibinfo {author}
  {\bibfnamefont {C.~D.}\ \bibnamefont {Batista}},\ and\ \bibinfo {author}
  {\bibfnamefont {S.-W.}\ \bibnamefont {Cheong}},\ }\bibfield  {title}
  {\bibinfo {title} {Topological defects as relics of emergent continuous
  symmetry and higgs condensation of disorder in ferroelectrics},\ }\href
  {https://doi.org/10.1038/nphys3142} {\bibfield  {journal} {\bibinfo
  {journal} {Nature Physics}\ }\textbf {\bibinfo {volume} {10}},\ \bibinfo
  {pages} {970} (\bibinfo {year} {2014})}\BibitemShut {NoStop}%
\end{thebibliography}%
\end{document}